\documentstyle[12pt]{article}
\begin{document}

\title{\large Spectral Variability and iron line emission in the ASCA Observations
\\ of the Seyfert 1 Galaxy NGC4051}

\author{Matteo {\sc Guainazzi}$^{1,2}$, 
Tatehiro {\sc Mihara}$^2$, \\  Chiko {\sc Otani}$^2$
 $\&$ Masaru {\sc Matsuoka}$^2$} 

\date{}

\maketitle

\begin{center}
\noindent
{\it $^1$Istituto di Fisica, Unit\`a G.I.F.C.O./C.N.R., \\
Via Archirafi 36, I-90123, Palermo} \\[0.25cm]
\noindent
{\it $^2$The Institute of Physical and Chemical Research, \\ 2--1 Hirosawa,
Wako, Saitama 350--01, Japan} \\[3.cm]
\end{center}
\noindent
Send correspondence to: \\
Matteo Guainazzi \\
SAX Scientific Data Center\\
c/o Nuova Telespazio \\
Via Corcolle 19 \\
I-00131 Roma
Italy \\
e-mail: matteo@napa.sdc.asi.it 

\vfill\eject

\begin{footnotesize}
\begin{center}
{\bf Abstract}
\end{center}
We present the results of an extensive analysis of the {\it ASCA} AO2 observation
of the Seyfert 1 galaxy NGC4051. The target exhibits broadband [0.5--10 keV]
variability by a factor $\sim 8$ on time scales $\sim 10^4 \ s$, with
a typical doubling time $\sim 500 \ s$. The spectrum is characterized
by a strong emission excess over the extrapolated power law
at energies $E \le 1 \ keV$. Absorption edges due to ionized oxygen species
OVII and OVIII are detected together with an emission-like feature at
$E \sim 0.93 \ keV$.
The OVII edge undergoes significant variability on a timescale as
low as $\sim 10^4 \ s$, whilst no contemporary variability of the OVIII
feature is detected. Typical variability time scales
place constraints on the location and the density of
the absorbing matter.
In the self-consistent hypothesis of a
high energy ($E \ge 2.3 \ keV$) power law reflected by an infinite
plane-parallel cold slab, a
photon
index change ($\Delta \Gamma = 0.4$) has also been observed; a natural explanation can be found
in the framework
of non-thermal Comptonization models.
The iron line is redshifted (centroid energy $E \sim
6.1 \ keV$) and broad ($\sigma > 0.2 \ keV$); multicomponent structure is
suggestive of emission from a relativistic accretion disk; however if the disk
is not ionized a contribution by a molecular torus or an iron
overabundance by a factor $\sim 1.5$ are required.
\\[1.cm]
\end{footnotesize}

\section{Introduction}

NGC4051 is a nearby ($z = 0.0023$) Seyfert 1 galaxy well known as a rapidly
variable X--ray source. It was extensively observed in the past
decade and both its spectral features and its variability deserved
great attention. An emission excess above
the extrapolation of a simple power law was discovered by the
{\it Einstein} (Urry {\it et al.} 1989) and {\it EXOSAT} (Lawrence
{\it et al.} 1985) experiments in the soft X-ray band.
Recent analysis of two pointed {\it ROSAT} observations
indicated the presence of a strong  absorption feature at energy
$\sim 0.85 \ keV$
(M$^c$Hardy {\it et al.} 1995); it was interpreted as
a blending of the photoionization absorption edges of OVII and OVIII.
Similar features have been commonly discovered in the
{\it ROSAT} spectra of Seyfert 1 galaxies
and explained
as due to the presence of highly ionized matter
surrounding the central source (the so called ``warm absorber'' model);
they have been confirmed by several {\it ASCA} targets 
(among the others
MCG-6-30-15, Fabian {\it et al.} 1994, NGC 3783, George, Turner $\&$
Netzer 1993, NGC3227, Ptak {\it et al.} 1994, NGC5548 and Mrk841, Otani
1996).
M$^c$Hardy {\it et al.} (1995) calculated self-consistently spectra emitted
by matter in photoionization equilibrium with the ionizing continuum,
assuming a spherical shell geometry and uniform
density; they yielded a value for the
ionization parameter $\xi \sim 300$ and a column density $N_{warm} \sim
8 \times 10^{22} \ cm^{-2}$, with an intrinsic energy index
$ \simeq 1.19$.

At higher energy, there was also much evidence of  deviation from the
simple power law trend. {\it Ginga} data have been widely analyzed and many
different models were claimed to be the best explanation for the observed
spectra. Matsuoka {\it et al.} (1990) supported an interpretation in terms of a partially covered power law
with solar abundance or a reflection model or a thermal bremsstrahlung
and a power law, whilst Kunieda {\it et al.} (1992) favoured
reprocessing by
cold blobs. The detection of an
iron K-fluorescence line by neutral or mildly ionized
matter provided further support to the hypothesis of
nuclear radiation reprocessing.

NGC4051 was observed by the {\it ASCA} satellite during the Performance Verification
phase (PV hereafter). The source fell into one of the SIS1 inter-chip gaps during it. The relevant
analysis was published by Mihara {\it et al.} (1994)
(hereafter Paper I) and their main conclusions can be summarized as follows.
The unprecedented combination of energy resolution and broadband coverage
allowed to discover two distinct absorption edges at energies $E_1 \simeq
0.74 \ keV$ and $E \simeq 0.92 \ keV$; they were interpreted as due
to highly ionized oxygen species OVII and OVIII, although the
energy of the latter was blueshifted ($\sim 6 \%$). That gave
the first direct measurement of warm absorbing gas in
NGC4051. A power law modified by a warm absorber could partly explain
the apparent soft excess, but the adding of a further thermal component
with $kT \sim 0.1 \ keV$ was required to remove it completely. The
best--fit underlying power-law photon index was $\simeq 1.88$. A narrow
emission line at energy consistent with K--fluorescence by neutral or
mildly ionized iron ($E \simeq 6.45 \ keV$) was also detected with an
upper limit on the FWHM of $\sim 460 \ eV$ and equivalent width $EW
\sim 170 \ eV$. 

In this paper we report the results of an extensive analysis of a second
{\it ASCA} observation performed during the AO2 phase, with a much longer
integration time than the PV phase. Timing and spectral analysis
and short term ($\le 2$ days) variability will be mainly addressed
and comparison with the PV phase results will be quoted only in such a
framework. In $\S$2 data reduction methods
and criteria are presented. In $\S$3 we will discuss briefly the broadband
variability and characterize it from a statistical point of view. $\S$4
will deal with the spectral analysis of the AO2 data, while in $\S$5 the
variability of the main grand spectral features observed will be studied
in detail. $\S$6 will be devoted to a discussion of the results, whilst
some concluding remarks will follow in $\S$7.

\section{Observation and data reduction}

The {\it ASCA} satellite (Tanaka, Inoue $\&$ Holt, 1994) observed NGC4051
during the AO2 phase between 6th July (15:28:34 UT) and 9th
July 1994 (10:40:27 UT), for a total exposure time of $\sim 1.55 \times 10^5 \ s$.
The Solid State Imaging Spectrometers (SIS hereafter) were operating in 1 CCD
mode and the data were collected in FAINT mode.
Dark Frame Error (DFE) and echo
corrections
(Otani $\&$ Dotani 1994)
were applied and the Charge Transfer Inefficiency (CTI) corrected Pulse Invariant SIS data used. Typical DFE values at the time when the observation was
performed were $\sim 1\div2$ ADU, corresponding to only $\simeq 7 \ eV$,
while the energy deficit for transfer was $\sim 2\div3 \times 10^{-5}$ per
transfer both for SIS0 and SIS1 chips used, which causes about 1$\%$
difference of the energy. Another source of gain uncertainties is the
Residual Dark frame Distribution (RDD); for 1-CCD mode such effect is smaller than
$5 \ eV$ even in the post-launch calibrations. The uncertenties on the energy
scale that can be associated with such corrections can be therefore estimated
as $\le 10 \ eV$ in the lowest energy {\it ASCA} SIS band and $\le 20 \ eV$
around 6 keV. A systematic error of such order has been taken into account
in the forthcoming results.
The Gas Imaging Spectrometers (GIS hereafter) were operating in PHA mode. Good time intervals
for the data analysis have been selected according to the criteria in
Table 1. They amount to
\begin{table}
\label{tab7}
%
%
\end{table}
a total integration time of $\sim 6.8 \times 10^4 \ s$ for both
detectors. This is more than twice the effective integration time of the
PV phase observation. Photons were extracted from a circular area
$\simeq 4.7'$ of radius for the GIS and $\simeq 3.2'$ for the
SIS. Data preparation and selection were performed via the XSELECT
package v.1.2; spectral fits
were carried out using the XSPEC program, v.8.5 (Arnaud {\it
et al.} 1991). \\

\section{Temporal behaviour}

Broadband light curves for the SIS0 and GIS2 detectors are
displayed in Figure 1 (binning time $\Delta t = 312 \ s$).
\begin{figure}
\label{fig1}
%
%
\end{figure}
Strong variability by a
factor up to $\sim 8$ on time scale $\le 10^4 \ s$ is
clear. Two flare--like structures
can be recognized between $10^4$ and $3 \times 10^4 \ s$ from the beginning
of the observation
and in the last $10^4 \ s$.
Events of very rapid variability with a doubling time of
$\sim$ few hundreds seconds are also relatively common.

X--ray variability on very short time scales has been reported by
many authors. Doubling time scales of few $\sim 10^3 \ s$ were found by
Marshall {\it  et al.} (1983) with the Imaging Proportional Counter on board
the {\it Einstein} satellite, by Lawrence {\it et al.} (1985) with the LE and
ME {\it EXOSAT} detectors and by Matsuoka {\it et al.} (1990) in the 1987 {\it Ginga}
observation. In Paper I, single
episodes of flux doubling on time scales of the
order of $\sim$ few hundreds of seconds were reported.
In order to characterize such variability from the statistical point of view,
an analysis of the lowest-doubling times has been performed
on the 100 s-binned light curves
of all the four detectors. The results for the SIS0 and GIS2
are shown in Figure 2.
\begin{figure}
\label{fig4}
%
%
\end{figure}
The $dN_{events}/d(\Delta t)$ distribution exhibits a flat behaviour with a
sudden decrease below some typical time scale $\tau$. Fit of the distribution
function of the form $A[1-e^{(-\Delta t/\tau)}]$,
where $A$ is a normalization constant equal to the asymptotic mean of the
distribution,
yielded the values: $\tau = 500 \pm 300 \ s$ for
SIS0 and $\tau = 400\pm300 \ s$ for GIS2 light curve, with a reduced $\chi^2_r
\sim 1$. We therefore conclude that NGC4051
presents broadband
statistically significant variability on timescale
as low as $\sim$ few hundreds of seconds. \\

\section{AO2 observation spectral analysis}

Previous X--ray observations of NGC4051 showed a very complicated and variable
spectral structure; in PV phase data a simple power law with
photoelectric absorption
by cold matter
with cosmic abundance was too a simple model to explain the various features
observed.

AO2 data confirm such outcomes in the light of a better S/N due to the
more than doubled effective integration time. Background subtraction has been
performed both with blank sky data files and with spectra obtained from a
source--free region in the same FOV of the source. The results
are
fully consistent and the ones obtained with the former
procedure are shown in the present paper.
Spectral analysis has been restricted to the bands [0.57-10 keV] for the
SIS and [0.7-10 keV] for the GIS detectors in order to avoid low
effective area energy channels and the systematic narrow-band spectral feature
observed in the SIS spectra around the K neutral oxygen absorption edge
($E \sim 0.54 \ keV$, Gendreau {\it et al.} 1995)
that could significantly affect the results in the
softer band. In Figure 3 the results of a fit with a
simple photoelectric absorbed
power law are shown.
\begin{figure}
\label{fig5}
%
%
\end{figure}
The $N_H$
value has been constrained
throughout the paper not to be lower than the Galactic value of $1.3
\times 10^{20} \ cm^{-3}$ (Dickey $\&$ Lockman 1990). The reduced $\chi^2$
is rather poor in both cases ($\chi^2_{SIS0+1}=1644/601$ d.o.f.,
$\chi^2_{GIS2+3}=1633/1304$ d.o.f.). The main contributions to the $\chi^2$ are
due to: a) a soft X--ray excess emission above the extrapolated power--law
below $E \sim 1 \ keV$; b) a broad emission
line feature centered at $E \simeq 6.1 \
keV$.

The average source
unabsorbed flux - calculated in the hypothesis of the
phenomenological ``L'' model (see $\S$3.1) - is $2.6 \times 10^{-11} \ erg \ cm^{-2}
\ s^{-1}$ in the 0.4--2.0 keV band and $2.4 \times 10^{-11} \ erg \ cm^{-2}
\ s^{-1}$ in the 2.0--10.0 keV band (average on the SIS spectra), corresponding to
source--frame luminosities of $0.58
\times 10^{42} \ erg \ s^{-1}$ and $0.54 \times 10^{42} \ erg \ s^{-1}$
at $z=0.0023$.
Here and hereafter $H_0 = 50 \ km \ s^{-1} \ Mpc^{-1}$ and $q = 0.5$ are
supposed.
Such results are consistent with the
2--10 keV flux range observed by previous experiments ($0.2 \div 5 \times
10^{-11} \ erg \ s^{-1} \ cm^{-2}$), including the PV {\it ASCA} observation
($2.2 \times 10^{-11} \ erg \ s^{-1} \ cm^{-2}$). \\

\subsection{Soft excess}

In Paper I, Mihara {\it et al.} (1994) identified absorption edges features due
to highly ionized oxygen species OVII and OVIII
in the SIS NGC4051 spectra of PV phase data. The adding
of such features to the
simple absorbed power--law best--fit model, although statistically justified
according to the F-test, is not able to remove the soft excess.
A complete removal of the soft excess via narrow features (absorption edges
and/or
narrow emission lines with Gaussian dispersion $\sigma$ fixed equal to 0)
requires 6 lines in the energy range $0.57 \le E \le 1.16 \ keV$
and no edges with
a $\chi^2 = 661/594$ d.o.f. Such explanation, although statistically
acceptable, is {\it de facto} simply a simulation of a continuum
soft excess emission due to the limited energy resolution available;
thus it will not be considered in the following.

Instead, adding a thermal continuum component
({\it i.e.} a blackbody) to the SIS spectrum produces
a dramatic decrease of the $\chi^2$ with $\Delta \chi^2 = 876$. However, even after this improvement
narrow absorption features are clearly visible in the spectrum and can be
modelled either with a couple of absorption edges and an emission line
(hereafter Model L) or
with three absorption edges (hereafter Model E), with comparable resulting
$\chi^2$ ($\chi^2_L = 624.4/598$ d.o.f., $\chi^2_E = 624.0/598$
d.o.f.). The
adding of each of the quoted features is justified
according to the F-test as shown in Table 2, that contains the
\begin{table}
\label{tab1}
%
%
\end{table}
best--fit parameters for the models described.
The fits in Table 2 have been performed on the spectral channels
in the range [0.57-5 keV], in order to avoid contamination by the
higher energy line structure.
Here and hereafter errors are at 90$\%$ level of confidence for one
interesting parameter ($\Delta \chi^2 = 2.71$, Lampton. Margon $\&$ Boyer 1976).
This underestimates the actual 90$\%$ confidence region for multi--parameter
fitting but nevertheless serves to provide an indication of the curvature of
the $\chi^2$ space near the best--fitting points.
Parameters with units of energies are quoted in
the source reference frame.
The energies of the softer
two absorption edges
($E_1 \simeq 0.74 \ keV$ and $E_2 \simeq 0.87 \ keV$) are fully consistent --
within the statistical
uncertainties -- with the K-shell photoionization energies of the oxygen
ion species OVII and OVIII and will be therefore associated with them in the
following. The depth of the OVII edge is almost the same in Model
L and E
($\tau \simeq 0.35$)
while the OVIII depth is model--dependent, ranging from $\tau^{(L)} \simeq
0.12$ to $\tau^{(E)} \simeq 0.31$. The narrow spectral features' properties are
not significantly
dependent on the shape of the continuum, as shown in Table 2
\begin{table}
\label{tab2}
%
%
\end{table}
where the best--fit parameters obtained when the blackbody is substituted
by a bremsstrahlung, power--law, or multitemperature disk blackbody
emission are shown. The temperature for optically thick emission is $kT = 180 \pm 20 \ eV$. The best-fit blackbody parameters correspond to a
typical emitting region size $R_{bb} \sim 1.3 \times 10^{10} \ cm$, while
the emission measure for optically thin thermal emission is $n_e^2 V \sim
2 \times 10^{63} \ cm$.

It's worthwhile to note that a standard scattering model ({\it i.e.} the
superposition of power-law spectra with the same indices seen through different
absorbing column densities) fails to reproduce the observed soft excess
($\chi^2_r \simeq 1.9$, with significant residuals in the whole $E \le 2 \ keV$
range).

If the average SIS spectrum
absorption features are
fit with only two absorption edges, as in Paper
I, their energies turn to be $E_1 = 0.738\pm0.013 \ keV$ and $E_2 =
0.927\pm0.017 \ keV$. This outcome is consistent with Paper I;
however, according to the higher S/N AO2 data,
$E_2$ is more likely to be a blending of a double structured feature.
If the model E (L) is applied to the PV
phase data, the resulting optical depths are $\tau_{OVII}=0.26^{+0.20}_{-0.11}$
($\tau_{OVII}=0.43^{+0.04}_{-0.03}$)
and $\tau_{OVIII}=0.13\pm0.10$ ($\tau_{OVIII} = 0.51\pm0.04$).

Some of these features lie within the useable energy bandpass of the GIS detectors.
The soft excess in the GIS2 average spectrum can be modelled with
a single absorption edge at $E = 0.94\pm0.05 \ keV$, with $\tau=0.34\pm0.15$ and
$\Gamma=1.88^{+0.04}_{-0.05}$ (F-test = 163); but fits with a comparable
statistical significance can be achieved by the adding of an emission line
at $E = 0.81 \pm 0.03 \ keV$ ($EW = 70^{+20}_{-40} \ eV$, $\Gamma =
1.86^{+0.05}_{-0.06}$) for GIS2 and $E=0.80^{+0.03}_{-0.06} \ eV$,
($EW = 60^{+20}_{-30} \ eV$, $\Gamma = 1.88^{+0.05}_{-0.06}$) for GIS3
or with a thermal continuum.

\subsection{Warm absorber model}

As already stated in $\S1$, absorption features from highly ionized oxygen have been widely observed
in the {\it ROSAT} and {\it ASCA} spectra of many Seyfert 1 galaxies and suggest
the presence of warm matter along the line of sight.
Models of absorption by ionized matter intervening along the line of sight (the so called ``warm absorber'') can be calculated and
compared with the available observational data, provided some simplyifying
assumption on the geometrical and physical properties of the absorbing
matter are made. We have used the photoionization
code {\sc Cloudy} (Ferland 1991) in order to calculate
self-consistent total spectra resulting
from the passage of ionizing continuum radiation through an optically-
and geometrically-thin spherical shell. A power--law shape has been assumed
for the incident spectrum in the whole energy range between 13.6 eV and 40 keV.
A complete characterization of the physical properties of the absorbing
matter is given once the parameter $\xi \equiv L/n_e R^2$ is provided,
where $L$ is the luminosity
of the ionizing continuum, $n_e$ the numerical electron density
and $R$ the distance between the central source and the inner edge of the
shell. In fact the physical properties of the absorber matter depend on the
ionizing flux $F \propto L/R^2 = n_e \xi$.
$n_e$ and $\xi$ are supposed uniform inside the shell,
while $R=10^{16} \ cm$ and $L = 10^{43} \ erg \ s^{-1}$.
The free
parameters in such models are the photon index $\Gamma$, the ionization
parameter $\xi$ and the
``warm'' hydrogen column density $N_W$. A grid table model has
been fitted to the SIS spectrum.
Figure 4 shows the best-fit data/model ratio,
\begin{figure}
\label{fig8}
%
%
\end{figure}
with $\chi^2=851/604$ d.o.f. The major contributions
to the $\chi^2$ are below $E \sim 1 \ keV$, where the warm
absorber features are present that should provide the bulk of information
on the ionization state of the matter. The best--fit parameters are:
$\Gamma \sim 2.12$, $log(N_W) \sim 22.27$ and $\xi \sim 110$. A prominent
emission-like feature is present at $E \sim 0.90 \ keV$ and the OVII edge
looks not to be well modelled. An acceptable $\chi^2$ can be obtained either
through the adding of a thermal component, like a blackbody ($\chi^2 = 694/602$
d.o.f.) or
through the adding of three emission lines at energies $E = 0.60 \pm 0.08 \ keV$,
$E = 0.69 \pm 0.08 \ keV$, $E = 0.93 \pm 0.07 \ keV$ ($\chi^2 = 660/600$ d.o.f.), but in both
cases the warm absorber matter best-fit parameters are not consistent with
the presence of both OVII and OVIII ions [for example, in the latter: $log (N_H) \ge 22.50$,
$\xi = 162 \pm 12$];
such solutions, although statistically acceptable, have to be
regarded only as purely mathematical.
Thus the model does not provide a satisfactory explanation of the data as
the approximations and assumptions appear to be too simple. A more detailed discussion
on the properties of the absorbing matter implied will follow in $\S$5,
in the light of the variability properties that it displays.

\subsection{Iron line emission}

Adding a Gaussian emission line to the average spectra of both SIS and
GIS detectors (in order to model the residuals around the Fe-K region)
improves significantly the $\chi^2$. The best--fit parameters
for a simple absorbed power--law and Gaussian line model are shown in
Table 3, where only
\begin{table}
\label{tab3}
%
%
\end{table}
the channels with $E > 2.3 \ keV$ have been included
in the fit in order to avoid contamination by the soft excess spectral
components. The line centroid energy:
$E_{SIS} = 6.3\pm0.2
\ keV$ and $E_{GIS} = 6.0 \pm 0.3 \ keV$;
the equivalent width is quite large compared
to the typical values observed in Seyfert 1 galaxies ($EW_{SIS} =
350^{+170}_{-150} \ eV$) when compared to {\it Ginga} results. The line has a clear
broad structure.
\begin{figure}
\label{fig6}
%
%
\end{figure}
Power--law photon index and
normalization are $\Gamma = 1.84^{+0.05}_{-0.03}$ and $N_{pl} =
7.3^{+0.5}_{-0.3} \times 10^{-3} \ photons \ keV \ s^{-1} \ cm^{-2}$
at 1 keV respectively.

A detailed analysis of the emission feature reveals
a much more complex
structure. $\chi^2$ is further improved by adding
two narrow ({\it i.e.} Gaussian dispersion $\sigma$ held fixed at 0)
Gaussian emission components with energies $E^{(2)} \simeq 6.4 \ keV$
and $E^{(3)} \simeq 5.5 \ keV$, albeit the significance of the third is
somewhat marginal in the SIS spectra ($\Delta \chi^2_{SIS} = 4$, $\Delta
\chi^2_{GIS} = 11$).
The best-fit parameters for such fits are again summarized in Table 3.
The presence of the new components affects slightly the centroid
energy of the main component that decreases by $\sim 2 \%$. 
We will refer
to such multi-component structure as ``emission complex'' in the following.

Iron K-fluorescence line is considered as a typical signature of the
reprocessing of the central source radiation by optically thick matter,
possibly in the form of an accretion disk around the putative black hole.
Such a scenario, investigated by the works of
Guilbert $\&$ Rees (1988) and Lightman $\&$ White (1988), predicts a
continuum ``bump'' at energies $E \ge 10 \ keV$. Since the appearance of
the line and the excess continuum emission are thought to be strongly
correlated, a power--law +
reflection by a ``cold'' ({\it i.e.} not ionized)
accretion disk + emission complex model has been applied both to
the SIS and GIS data.
The inclination of the disk has been fixed to $\theta = 30^{\circ}$ (see
below), but the dependency of the following results on this parameter is
completely negligible. 
The only free parameter of the reflected spectrum is then the solid
angle subtended by the reflecting matter $\Omega/2 \pi$.
Adding of this new degree of freedom is not justified
according to the F-test ($F_{GIS} = 1.0$, $F_{SIS}=0.1$); only a very loose
constraint on the solid angle
($ \Omega/2 \pi \le 7$) is yielded 
if the model is applied on the spectra of the four detectors simultaneously.
In the hypothesis of reflection by a plane--parallel slab ($\Omega/2 \pi$
held fixed
to 1), the best--fit parameters for the main emission complex component
are: $E^{(1)} = 6.1\pm0.3 \ keV$, $\sigma^{(1)} = 0.7^{+0.7}_{-0.5} \ eV$, 
$EW^{(1)} = 220^{+290}_{-150} \ eV$, with an intrinsic photon index
$\Gamma = 1.88^{+0.03}_{-0.04}$. The additional complex components have:
$E^{(2)} = 6.44\pm0.05 \ keV$, $EW^{(2)} = 60\pm30$,
$E^{(3)} = 5.50^{+0.10}_{-0.19} \ keV$ and $EW^{(3)} = 16^{+26}_{-15} \ eV$
respectively.
The contour plot for the centroid energy vs. the Gaussian dispersion
of the main component is
shown in
the upper panel of Figure 5; at 90$\%$ level of confidence the Gaussian dispersion
is $\ge 0.2 \ keV$.

The total equivalent width of the emission complex is much higher than the
average observed by {\it Ginga} in the same target ($\langle EW \rangle_{NGC4051} =
140\pm70 \ eV$, Nandra $\&$ Pounds 1994), where a single ``narrow'' line was supposed. It is also
much higher than the mean of the observed EW distribution in the complete
{\it Ginga} Seyfert 1 sample ($\langle EW \rangle_{Ginga} = 140 \pm 20 \ eV$,
Nandra $\&$ Pounds 1994). However several Seyfert 1 observed by {\it ASCA}
showed clearly a broad fluorescence iron line: IC4329A ($\sigma_{99\%} \ge 0.15 \ keV$,
Mushotzky {\it et al.} 1995, Cappi {\it et al.} 1996), NGC5548 ($\sigma_{99\%} \ge
0.1 \ keV$, Mushotzky {\it et al.} 1995), MCG-6-30-15 ($\sigma_{90\%} = 0.16^{+0.10}_{-0.06} \ keV$, Fabian {\it et al.} 1994).
Complex and broad emission structures are to be expected if the emission originates in an
accretion disk.
As stated by many authors (Fabian {\it et al.} 1989, Matt, Perola $\&$ Piro
1991), the Doppler shifting, gravitational reddening and
relativistic beaming
are effective in altering the Gaussian profile of an emission line and
producing a typical doubled--horned structure.
Compelling evidence of such kind of
structure in the {\it ASCA} observation of the Seyfert
1 MGC-6-30-15 (Tanaka {\it et al.} 1995) and of the intermediate Seyfert NGC4151
(Yaqoob {\it et al.} 1995)
have been recently discovered,
giving significant support for such a scenario.

In order to check quantitatively
the disk hypothesis, line emission from a relativistic accretion
disk model by Fabian {\it et al.} (1989) has been applied to the [2.3-10 keV]
spectra of all detectors simultaneously. The free parameters are the inner
($R_i$)
and outer ($R_o$) radii of the line emitting region in units of Schwarzschild
radii $R_S$ from the center and the inclination angle $\theta$ between the
normal to the plane of the disk and the line of sight. The energy of the
line centroid $E_{\alpha}$ has been held fixed to 6.4 (cold disk) and
6.7 (ionized disk). The emerging profile is also affected by the radial
dependence of the emissivity, that in such a model is parameterized with a
function $R^{-q}$.
The underlying continuum is supposed to be a power--law with reflection by
a plane-parallel slab and the disk-line inclination tied to the
reflecting slab inclination.
The best--fit parameters are listed in Table 4. The
observed line is consistent either with a cold disk at moderate inclination
($\theta \sim 30^{\circ}$) or with an almost face-on ionized disk. $R_i$
is $\sim 10 R_S$, while only in the cold disk case $R_o$ can be constrained
to a value $\sim 30 R_S$. In the cold case the derived EW is higher
than the maximum allowed by presently available models with solar abundance,
by a factor $\sim 1.5$.

We stress that no absorption edge by neutral or ionized iron has
been detected in the average AO2 SIS spectra. Upper limits on the optical
depth for some trial energies have been: $\tau_{FeI}(E = 7.11 keV) <
0.06$, $\tau_{FeXX}(E = 8 \ keV) < 0.19$, $\tau_{FeXXIV} (E = 8.80 \ keV)
< 0.05$. \\

\section{Spectral variability}

NGC4051 is known to be a variable X--ray source. Both {\it EXOSAT} (Lawrence
{\it et al.} 1985) and {\it Ginga} (Matsuoka {\it et al.} 1990) observations
showed a hardening of the spectrum for decreasing flux; in both cases it was
described in terms of decrease of the spectral photon index.
On the other hand, Kunieda {\it
et al.} (1992) analyzed successfully the spectrum of the May 1988
{\it Ginga} observation
in terms of reprocessing of the nuclear radiation by blobs of matter
($N_{blob} \sim 10^{24.5} \ cm^{-2}$). Significant spectral changes
accompanied flux variations by a factor of 2--3 within 1000 s in the
softer {\it Ginga} band (2.3--6.4 keV), but they did not strictly follow
the luminosity trend; instead, it was possible to distinguish three
different phases, characterized by a constant hard
({\it i.e.} 8.7-20.9 keV) flux
with hard spectrum, by a low energy flare with soft spectrum and by a
soft spectrum with almost synchronized flux variability respectively.
Such complex spectral changes could be
successfully explained by changes of the intrinsic
luminosity and the blob number along the line of sight. Fiore {\it et al.}
(1992a) re-analyzed all the {\it Ginga} NGC4051 observations in the framework
of two-component models (power--law + reprocessing) and 
affirmed a simultaneous variation of both. A 99$\%$ level of confidence
variability of the iron line correlated with that of the underlying
continuum down to time scales as short as few thousands seconds gave further
support to such hypothesis.

 The {\it ASCA} AO2 observation data provide a lot of information about the
variability of the source and  allow to deliver new insight on this topic. In
the following the soft excess and
intrinsic power--law photon index variability will be
mainly addressed, both via
hardness ratio analysis and via direct comparision of spectral fits performed on
different intensity phases.

\subsection{Hardness ratio analysis}

The broadband spectrum of the NGC4051 AO2 {\it ASCA} observation shows
flux-dependent changes at both low and high energies. The PHA ratio of the
average SIS0 spectra corresponding to a High and Low intensity State is
shown in Figure 6. An operative definition of the procedure
\begin{figure}
\label{fig14}
%
%
\end{figure}
followed to select High and Low State photons will be given in $\S$5.2.
The High State spectrum is harder for $E \ge 2 \ keV$ and
$E \le 1 \ keV$. 

A useful tool for characterizing the spectral variability of X--ray
sources is the study of Softness
Ratio ($SR$ hereafter). The energy
bands have been chosen in such a way
they are representative of the behaviour of different
spectral components; according to Netzer, Turner $\&$ George (1994)
the most suitable choice is
the bands: [0.7-1.3 keV] and [2.5-5 keV]. The SR is therefore defined as:
\begin{equation}
\label{eqn2}
SR = \frac{\mbox{CR [0.7-1.3 keV]}}{\mbox{CR [2.5-5.0 keV]}}
\end{equation}
Such choice allows to study the differential variability
of the spectral components. 
Warm absorber should predominantly affect the flux in the softer band.
The low energy cut--off of the
harder band has been chosen in order to avoid contamination by the
iron line emission complex.

In Figure 7 the SIS0+1 light curve in the two bands and the SR light
\begin{figure}
\label{fig9}
%
%
\end{figure}
curve are shown for a binning time $\Delta t = 500 \ s$.
The $\chi^2$ for
constant hypothesis of the $SR$ light curve is $\chi^2_{SR} = 341.2/61$ d.o.f.
The $SR$ undergoes a significant variability, getting
lower for increasing flux and higher for decreasing flux. When the source
flux approaches its minimum throughout the observation ($1.2 \times 10^5 \
s$ after the start), the $SR$ doubles within a time scale
$\sim 10^4 \ s$. A decrease of comparable order of magnitude seems to
be associated with the sudden doubling of flux within $\sim$ few thousands of
second after $\sim 10^4 \ s$ from the start.

The determination of the correct variability timescale is of the utmost
importance in order to derive meaningful constraints on the geometry and
physical processes involved in Active Galactic Nuclei
(AGN hereafter).
The $SR$ dependency
on the total count rate has then been studied in detail. In Figure 8
\begin{figure}
\label{fig10}
%
%
\end{figure}
the $SR$ is plotted vs. the total [0.7-5 keV] count rate
(binning time $\Delta t = 5000$).
The $\chi^2$ for constant hypothesis is $\chi^2 = 259/29$ d.o.f.,
corresponding to a chance occurrence likelihood $P< 0.1\%$.
Also the linear correlation coefficient is rather high ($r_{29}=0.73\pm0.05$),
again corresponding to a chance occurrence likelihood $< 0.1 \%$.
A careful visual inspection at the
$SR$ light curve in Figure 7
would suggest the highest variability to be concentrated
in the $\sim 2 \times 10^4 \ s$ time interval
when the source flux is minimum. The
same analysis as above has then been repeated
excluding it from the light curves; however the strong
$SR$ vs. $CR$ correlation has been confirmed
($\chi^2 = 131/24$ d.o.f., $r_{24}=0.64\pm0.10$, $P<0.1\%$ for both parameters).
Similar results
are obtained for binning times
$\Delta t = 500 \ s$ ($\chi^2 = 320/61$ d.o.f., $r=0.73\pm0.05$) and 
$\Delta t = 25000 \ s$ ($\chi^2 = 151/6$ d.o.f., $r=0.95\pm0.03$, $P \simeq
0.5\%$). The binning time is obviously limited by the total elapsed time
of the observation ($T_{elaps} \sim 1.5 \times 10^5 \ s$), and no significant
piece of information can be derived for $\Delta t > 25000 \ s$. Anyway,
there is compelling evidence of an increase of the $SR$ when the source flux
increases on the whole two order of magnitude range between $\Delta t \sim
10^2 \ s$ and $\Delta t \sim 10^4 \ s$.
That evidence can naturally be explained
in terms of a lower hard flux or weaker absorption by ionized matter
when the source is more intense or any combination of the two. In order to
disentangle
the two effects (or others that might be present) a
detailed time-resolved analysis of the various spectral components is needed.

\subsection{Spectral analysis: soft excess}

The high effective integration time allows a detailed time--resolved analysis
of the main grand spectral features discovered in the NGC4051
average spectrum. For
such purpose, the whole observation has been divided into 11 contiguous
segments according to the following procedure: a $\Delta t = 100 \ s$ light
curve has been extracted and all the consecutive bins whose count rate was
higher then the total mean count rate have been gathered together
(contamination not higher than $\sim 12\%$ has been allowed).
The same procedure has been followed
for bins whose count rate was lower than the
total mean count rate. The log of
the resulting subdivisions is shown in Table 5. Each of such
\begin{table}
\label{tab4}
%
%
\end{table} 
subdivision will be referred to as {\it orbit} in the following, although
they have no relation with the real orbits of the {\it ASCA} satellite.
The span of each orbit is never less than $\sim 7 \times 10^3 \ s$
except in the case of the 3rd orbit whose time span is $\sim 4 \times 10^3 \ s$.
The count
rate for each orbit is
in the range $60\%-144\%$ of the total mean
count rate. The minimum flux episode is included in the 10th orbit, the
flare--like structures in the 2nd and 11st.

The best--fit 
optical depths for OVII and OVIII edges when Model L is applied to the
combined SIS0+1 spectrum of each orbit are shown in
5th and 6th columns of Table 5.
In such fits the energies of the edges are frozen
to their best--fit values in the average spectral fit
due to low S/N.
The 10th orbit OVII optical depth
is significantly higher than the mean. For the sake of a deeper investigation,
we have gathered together the orbits
1+3+5+7+9+11 (hereafter referred to as High State, HS)
and the orbits 2+4+6+8+10 (Low State, LS) and applied Model E and L to
the SIS HS and LS spectra. The results are summarized in Table 6.
\begin{table}
\label{tab6}
%
%
\end{table}
The OVII optical depth undergoes a significant
($90\%$ level of confidence) change
by a factor $\sim 50\%$, the edge energies remaining consistent within
the quoted statistical uncertenties. No significant variation
is observed either for energy or optical depth of the OVIII edge.
A further detailed analysis has been performed splitting the 10th
orbit from the rest of Low State,
in order to check the significance of the outcomes summarized
in Table 5. We will
refer hereafter to the Low State phase when the 10th orbit is subtracted as
LS$^{\star}$. A
99$\%$ level of confidence contour plot for the OVII
edge energy vs. optical depth when the L model is applied to HS,
LS$^{\star}$
and 10th orbit is plotted in Figure 9. The 10th orbit and HS
\begin{figure}
\label{fig13}
%
%
\end{figure}
optical depths are different at more than $99\%$ level of confidence, whilst
the LS$^{\star}$ optical depth is not consistent with the other two phases'
at only 90$\%$ level of confidence. At this latter level, however, it can be
stated that a smooth trend of decreasing of OVII optical depth
with increasing flux is observed
on timescales as low as $10^4 \ s$, and then its variability
is not limited at the lowest flux interval. No significant variability
within the quoted statistical uncertainties is instead detected for the
\begin{figure}
\label{fig12}
%
%
\end{figure}
OVIII absorption edge (see Figure 10).

The spectral shape in the lowest energy {\it ASCA} band is very complex;
the observed variability of the narrow-band features
might be due to the interplay between the ionized absorption and the continuum
soft excess component. It must be remembered however that the ``L'' model
comprises
self-consistently all the spectral components required to remove the soft
excess. Moreover, the observed variability patterns are almost independent
of the detailed shape of the soft continuum component used. In Figure 11
the optical depth vs. energy contour plots are shown when a bremsstrahlung
or power-law models replace the blackbody. Analogous conclusions
can be derived about the edges' variability. The continuum models in Figure
11 are likely to be less physically meaningful explanations of the observed
soft excess emission; but the lack of any significant dependency of the
narrow-band feature results on the underlying continuum spectral shape gives
strong support to their reality.

\subsection{Spectral analysis: power--law photon index}

The 7th row of Table 5 shows the best--fit value that
the power--law photon index $\Gamma$ assumes
when the L model is applied. It is significantly variable
($\chi^2$ for constant hypothesis 28/10 d.o.f.) and appears to be correlated
with the total flux ($r_{10} = 0.8\pm0.3$).
If the soft excess is efficiently removed by models
``E'' or ``L'', this should be an indication of a real variation of the
observed underlying continuum photon index.

Assessing unambiguously an intrinsic variation of the spectral index is
however a very difficult task if the {\it ASCA} data alone are available.
Any variation in the relative normalization of the direct and reflected
components might produce an apparent change in the steepness of the spectrum.
If, as the outcomes in $\S4.3$ might suggest, the iron line comes from a
relativistic accretion disk, its broad and skewed structure might give a
significant contribution to the photon emission in the 4-8 keV range. Last,
but not least, absorption by ionized matter might affect the high energy
spectrum also above $2.3 \ keV$. The limited {\it ASCA} high energy bandwidth
does not give a good statistics above $\sim 8 \ keV$ and it is thus
difficult to constrain the spectral shape in the hard tail. Having all
such problems in mind, we searched for index changes, provided a self-consistent
physical model with some simplifying assumptions is assumed, in order to
reduce the number of degrees of freedom; namely:

\begin{itemize}
\item[{\it i)}] we assumed a power law spectrum + reflection by a plane-parallel
infinite slab + iron emission line complex.
\item[{\it ii)}] The fainter
components' parameters have been held fixed to their average best-fit value
(see $\S4.3$)
\item[{\it iii)}] neither the absorbing medium nor the continuum soft excess
component have been supposed to affect substantially the spectral shape for $E \ge 2 \ keV$. That
implies the numerical density of the intervening matter $\le 10^{22} \
cm^{-3}$ and the temperature of the blackbody of the order at most few hundreds eV.
Both conditions are fulfilled in the {\it ASCA} data according to the low energy
bandpass outcomes (see $\S6.2$)
\end{itemize}

Such a model has been fitted to the SIS+GIS spectra obtained
when the following orbits are gathered together: 8+10 (1st quartile),
2+6 (2nd quartile), 4+5+7+9 (3rd quartile), 1+3+11 (4th quartile).
The relative normalization between the direct and reflected components
has been kept constant ($\Omega/2 \pi = 1$) or allowed to vary inversely
to the flux; such two cases simulate an immediate or delayed response
of the reprocessing to nuclear flux changes respectively. In the latter
scenario, the intrinsic width of the main emission complex component
has been held fixed to its average best-fit value in order to further
reduce the available degress of freedom.
The best--fit parameters are listed in Table 7 and 8. $\Gamma$ shows the same
qualitative trend in both cases.
In Figure 12 a plot of the $\Gamma_{2.3-10 keV}$ vs. the [2-10 keV]
luminosity
\begin{figure}
\label{fig11}
%
%
\end{figure}
is shown in the constant normalization ratio case. The figure includes also two measures
when the same model is applied
to the spectra of the HS and LS PV phase data (the definition of
the intensity phase in PV data is analogous to the AO2 case).
\begin{table}
\label{tab5}
%
%
\end{table}
If all the six experimental points are taken into account the
index is variable ($\chi^2$ for constant hypothesis 33/5 d.o.f.) and
correlated with the [2--10 keV] luminosity ($r_6 = 0.88\pm0.08$, the
lower limit corresponding to a likelihood of casual occurrence
$P < 6 \%$). Such correlation is valid also if only the four
experimental points relative to the AO2 data are considered ($\chi^2$
for constant hypothesis 31/3 d.o.f., $r_4 = 0.98\pm0.04$, $P < 25\%$).
In Figure 12 it is straightforward to see that the PV phase measuraments are not
aligned  exactly with the same straight line corresponding to the AO2 measures. A secular
pattern of variability could therefore be superimposed on the short
timescale correlation between spectral index and luminosity. If the
PV phase experimental points are arbitrarily shifted rightward by an amount
$\sim 20 \%$, the alignment is qualitatively much better and
the linear correlation coefficient $\Gamma$ vs. $L$ turns to be consistent
with 1 even for all the six measures ($r_6 = 0.97\pm0.04$);
we remind here that the average [2-10 keV] flux of the AO2
observation is just $20\%$ higher then the flux of the PV phase data
in the same energy range. Hysteresis phenomena in the $\Gamma$ vs. $L$
relation should be expected if the continuum is produced in a pair
dominated Comptonized non-thermal plasma (Yaqoob 1992). If we take the
extreme points in Figure 12 as representative of the low and high state of
the source, the pivot point for spectral change is $E_p \simeq 20 \ keV$.

Next question to be addressed  is whether the observed changes in the
spectral index are really due to changes in the slope of the
intrinsic power--law.
As stated in section 4.3, the presence of the iron line
is supposed to be correlated to a ``bump'' of
continuum emission at energies
$E \ge 10 \ keV$. Differences in the observed photon index could
therefore be due to different amount of reflected flux, with
the stronger reflection producing a flatter spectrum. In order
to check this hypothesis we have first calculated the ratio between
the photon indices in the [2.3-5.0 keV] and [5.0-10.0
keV], the former band likely to be unaffected by any reprocessed
component;
the measure of such ratio for all the quartiles is reported in 6th
column of Table 7. Although a slight trend of decreasing
of such ratio with increasing luminosity can be noticed, all the
values are consistent with 1. As a further check, fits
have been performed when the $\Omega/2 \pi$ parameter in the reflection
model is left free.
The adding of this further degree of freedom
is not statistically justified for any of the AO2 quartiles
(see 7th column in Table 7) and
in neither the HS nor the LS phase in PV data.
 
We conclude
therefore that the $\Delta \Gamma \sim 0.4$ variation observed in the
{\it ASCA} data of NGC4051 might be interpreted as due to an intrinsic
index change in the framework of a self-consistent physical model that
takes into account all the spectral components in the reflecting
accretion disk scenario. Such outcomes partly confirms the results of
Matsuoka {\it et al.} (1990). 

\section{Discussion}

\subsection{Broadband variability}

The outcomes presented contribute to shed new light on the spectral properties
of the X--ray emission from AGNs and allow to set constraints
on the geometrical and physical properties of the processes
involved.

The main source for the huge amount of power emitted by the nuclear region
of AGN is the conversion of gravitational
to radiative energy in matter falling onto a supermassive
central accretor. NGC4051 is well known as one of the most variable
extragalactic sources; {\it ASCA} observations revealed flux changes by a factor of
8 within typical timescale $\sim 10^4 \ s$ and doubling time $\sim$ few
hundreds seconds.
The {\it ROSAT} light curve in the softer [0.1--2.4 keV] range 
displayed variations by a factor $\sim 10$ within $\sim 4
\times 10^4 \ s$ (M$^c$Hardy {\it al.}  1995).
Light crossing time arguments yield an upper limit for the
mass of the central object $M \le 10^7 \ M_{\odot}$ if the size of the
emission region is as large as 5 Schwarzschild radii. We stress that
previous estimates based on the correlation between the iron line emission
strength and the underlying continuum intensity - that yielded values by a
factor of 5 lower - are not confirmed by the {\it ASCA} data, since no
flux-correlated variability
of the iron line could be firmly assessed within the 2 days AO2 observations
(cfr. Table 7).
The required efficiency in converting accreting matter to luminosity
(Fabian $\&$ Rees
1979) is $\ge 10^{-3}$, consistent with accretion onto a Schwarzschild
black hole.

\subsection{Identification of the third narrow low--energy spectral
feature}

The AO2 spectra have provided further confirmations of the presence of narrow
features due to absorption by highly ionized matter.
Better AO2 data S/N ratio has allowed to recognize
the absorption edge discovered in PV phase data at energy $E \sim 0.92
\ keV$ as probably due to blending of two different features. Two
edges at energies $E \sim 0.74 \ keV$ and $E \sim 0.86 \ keV$ have then
been detected in the AO2 SIS spectrum, whose energies are fully consistent
with the K-photoionization energies of the ionized oxygen species OVII
and OVIII. A third narrow feature is statistically required to model
the spectrum, but it is not possible to distinguish between a further
absorption edge or an emission line
from a purely statistical point of view. A correct identification of
the nature of this third feature is of primary importance, also
because the
OVIII absorption edge optical depth is strongly model dependent. We
will therefore address this topic in this section.

Provided this third feature is an absorption edge, it could be a blending of
the K-photoionization edges of NeIII and/or NeIV ($937 \le E \le 971 \ eV$)
or a blending of L-photoionization edges of FeVII$\div$X ($930 \le E \le
1000 \ eV$). If such species are to be located in the same physical region
as OVII and OVIII, the ionization structure and temperature of the
corresponding absorbing matter
has to be the same.
In the ``E'' model the typical
values of oxygen optical dephts are $\tau_{OVII} \sim 0.35$ and
$\tau_{OVIII} \sim 0.12$, corresponding to hydrogen equivalent column density
of $N_H(OVII) \sim 1.6 \times 10^{21} \ cm^{-2}$ and $N_H(OVIII) \sim
1.3 \times 10^{21} \ cm^{-2}$.
The range of ionization
parameters that allows an abundance ratio $[OVII]/[OVIII] \sim 1$
for a cloud of gas in photoionization equilibrium
is relatively
narrow ($\xi = 50 \div 90 \ erg \ cm \ s^{-1}$, Kallmann $\&$ McCray 1982),
with typical plasma temperature $\sim 2 \times 10^5 \ K$.
On the other hand, the presence of
NeIII/NeIV or FeVII$\div$X species implies $\xi$ not to be higher than
25.
Moreover the equivalent hydrogen FeVII$\div$X column density required would be
$\sim 8 \times 10^{22} \ cm^{-2}$.
One could suppose
the oxygen features to be produced in a physically different region than
the third absorption feature; such complication
looks however unnecessary, although it cannot be
ruled out {\it a priori}.

A plasma in photoionization equilibrium with an ionizing continuum is
expected to emit a
rich spectrum of X--ray lines as the result of collisional excitation,
inner shell fluorescence and recombination.
Line-like features with
typical $EW \sim 30-100 \ eV$
were detected in the {\it Einstein SSS} spectra
of several Seyfert 1 galaxies (Turner {\it et al.} 1991) and suggested to
originate in a hot ($T \sim 10^7 \ K$) collisionally excited plasma.
However subsequently Netzer (1993) pointed out that photoionized gas might
be responsible for them.
One of the most promising candidates in the range of the observed NGC4051
line(s) ($\lambda \ 13.0-13.5  \AA$) is L-shell iron line-complex which can
cover the whole range between 0.7 and 1.6 keV (Band {\it et al.} 1990,
Liedhal {\it et al.} 1990).
However, iron L-shell lines should dominate the spectrum for values
of the ionization parameters $U \equiv F_{ion}/N_H c > 10$, where $F_{ion}$
is the ionizing continuum flux and $c$ is the speed of light. That corresponds
to $\xi \ge 100$, slightly higher than the one inferred by the oxygen ions
abundance. Line--like features with energy $\sim 900 \ eV$ could therefore be
likely produced by OVII and OVIII continua or NeVII/IX $L_{\alpha}$
lines. 

\subsection{Warm absorber variability}

The presence of absorption edges due to highly ionized oxygen advocates
for a distribution of matter along the line of sight
whose ionization
structure is -- partly or totally -- sustained by the radiation emitted
by the AGN central source.
In such physical conditions, the
state of the matter can be parametrized through the electronic temperature
$T$ and an ionization parameter like $\xi$
or $U$. 
Krolik, McKee $\&$ Tarter (1981) showed
that in conditions of thermal and ionization equilibrium matter should be
thermally stable only in two phases: a cold ($T \le 10^5 \ K$) phase with
lower ionization states and a hot ($T \ge 10^7 \ K$) phase with higher
ionization states. However, a recent paper by Reynolds $\&$ Fabian (1994),
stated the existence of a possible intermediate stable state with typical
$T \sim 10^5 \ K$ and $\xi \sim 10^{1.5-2}$, which they proposed to be
a good candidate for the warm absorber. Such state
would be consistent with the co-existence of OVII and OVIII species;
the range of allowed ionization parameters is relatively narrow
[$\Delta \Xi/\Xi \sim 0.1$, where $\Xi \equiv (2 \times 10^4 K/T) \xi$] and
that allows to study the warm absorber variability in terms of fluctuations
around a stable state where both OVI and OIX numerical density is
negiglible if compared to the prevailing oxygen ions.

The detected variability of absorption features in the {\it ASCA} AO2 observation
of NGC4051 allows to put some constraints on the geometrical and physical
properties of the absorbing matter.
At first we suppose
density and ionization parameter of the absorbing matter to be uniform
(one--zone hypothesis) and the ionization structure to be driven predominantly
by the photoionization equilibrium. In such hypothesis the relevant time scales
for each ionized species are the radiative recombination and the
photoionization time scale to be compared with the typical variability
time scale $t_{var} \sim 10^4 \ s$. The most natural explanation for the
decrease in the OVII edge optical depth with increasing flux is the increasing
in the amount of photoionization processes OVII$\rightarrow$OVIII.
The typical ionization timescale can be expressed as ({\it e.g.} Otani 1996):
\begin{equation}
\label{eqn9}
t^{OVII}_{ph} \sim 80 \left( \frac{100}{\xi} \right) \left( \frac{10^9 \ cm^{-3}}{n_{OVII}}
\right) \left( \frac{E_{edge}}{1 \ keV} \right) \left(
\frac{10^{-19} \ cm^{-2}}{\sigma_{ph}(E_{edge})} \right) \ s
\end{equation}
$$
\simeq 4 \times 10^{10}/n_{OVII} \ s
$$
where $n_{OVII}$ is the numerical hydrogen equivalent 
density of OVII and $\sigma_{ph}
(E_{edge})$ is the photoionization cross--section at the absorption edge
energy. From the condition $10^4 \ s \ge t_{var} \ge t_{ph}$ it is possible
to set $n_{OVII} \ge 4 \times 10^6 \ cm^{-3}$, that implies a lower limit for
the density of the absorbing matter $n \ge 2.1 \times 10^{10} \ cm^{-3}$ assuming
cosmic oxygen abundances.
$t_{ph}$ can be also expressed as a function of physical parameters of the
warm absorber:
\begin{equation}
\label{eqn10}
t^{OVII}_{ph} \sim 20 R^2_{16} L^{-1}_{43} \ s
\end{equation}
where $R_{16}$ is the distance of the warm absorber from the nuclear source
in units of $10^{16} \ cm$ and $L_{43}$ the X--ray luminosity in units of
$10^{43} \ erg \ s^{-1}$. The variability constraint implies therefore
$R_{16} \le 7$.
If the change of the photoionization opacity produces
a change in the column density $\Delta N_H \sim \Delta R \ n$, it follows
$\Delta R \le 3 \times 10^{15} \ cm \ll R$.
That should be suggestive either of geometrically thin or -- more likely --
of a quite small filling factor for the distribution of the absorbing matter.

A change in the column density of the absorbing OVII should be accompanied
by an opposite change in the OVIII column density. However, in the {\it ASCA}
NGC4051 data there is no clear evidence of such change.
The relevant time scales involved for the OVIII balance are the recombination
timescale $t_{rec}$ for the OVIII$\rightarrow$OVII process and the
photoionization timescale $t^{OVIII}_{ph}$. The former is expressed by ({\it e.g.} Otani 1996):
\begin{equation}
t_{rec} \sim \frac{1}{\alpha_{rec} n_{OVIII}}
\end{equation}
$$
\simeq 2 \times 10^7 \frac{T^{0.76}}{n_{OVIII}} \sim 2 \times 10^{11}/n_{OVIII}
\ s
$$

In the framework of the ``L'' model, $n_{OVIII} \sim 3 n_{OVII}$ and
$t_{rec} \sim t_{ph}^{OVII}$. The condition of ionization equilibrium
for OVII is therefore likely to be fulfilled. In the meanwhile, the OVIII
photoionization timescale is given by:
\begin{equation}
t^{OVIII}_{ph} \sim 1.3 \times 10^{11}/n_{OVIII} \ s
\end{equation}
according to equation~\ref{eqn9}. Then $t_{rec} \sim t^{OVIII}_{ph}$ and that
could contribute to keep the OVIII species constant.
 
Similar evidence of such uncorrelated variation of ionized
oxygen edges has been observed
in the {\it ASCA} observation of MGC-6-30-15 (Reynolds {\it et al.} 1995,
Otani {\it et al.} 1996), where a
significant change in the absorption optical depth of OVIII was
detected with no appreciable change in OVII. 
A stratification of the absorbing medium
was claimed as a possible interpretation.
A recent paper by Krolik $\&$ Kriss (1995) has pointed out that
there is no reason {\it a priori} for imposing either the thermal equilibrium
or the ionization balance; whether or not they are fulfilled depends strongly
on the detailed ionization structure, on the mean luminosity and variability
features of each AGN. The presence of resonance scattering lines can moreover greatly
affect both the energy and the optical depth of the observed absorption
features, leading to an incorrect interpretation of absorbing medium
properties. Taking that into account can lead, for instance, to a dramatically
different
interpretation of the low--energy spectral variability in MGC-6-30-15,
suggesting a fixed column density and ionization equlibrium scenario.
Some caution must then be employed in the interpretation of the NGC4051 quoted
results and only high resolution spectroscopy in the soft X-ray band will
be able to solve all the puzzling questions still open.

\subsection{Photon index variability}

{\it ASCA} detected a significant ($\Delta \Gamma \sim 0.4$) and robust change
of the photon index for a factor of 4 flux variation. A still
higher variability ($\Delta \Gamma \sim 0.7$)
had been found in {\it Ginga} data (Matsuoka {\it et al.}
1989), although a possible explanation in terms of variable absorption
could account for the data equally well (Fiore {\it et al.} 1992b). The
broadband {\it ASCA} coverage and the improved high energy resolution have
allowed to conclude that at least part of such variability can be explained
in terms of intrinsic photon index variation.

An analogous change in the power--law slope was found by Leighly {\it et
al.} (1996) in the {\it ASCA} observation of the Narrow Line Seyfert Galaxy
Mrk766, that showed a $\Delta \Gamma=0.4$ for a flux variation by a factor of
$\sim 2$ on typical timescales $\sim$ few thousands of seconds.
A renewed interest has thus recently arisen about such topic.

The most successful model for the continuum production in AGN is the
Comptonization of UV or soft X--ray seed photons,
which probably originate in an accretion disk, by a distribution of
relativistic electrons. Depending on the nature of the injected electron
distribution such models can be {\it thermal} or {\it nonthermal}. A
spherical geometry for the interaction region is assumed with radius
$R$ and luminosity $L$, but a fairly independent description by the
details of the involved geometry can be achieved if the dimensionless
luminosity to size ratio, or {\it compactness parameter} $\ell$ is introduced
(Cavaliere $\&$ Morrison 1980, Guilbert, Fabian $\&$ Rees 1983),
defined as: $\ell \equiv \frac{L}{R} \frac{\sigma_T}{m_e c^3}$, where
$\sigma_T$ is the Thompson scattering cross section. A further ingredient
in such models is the $\gamma$\--ray absorption via electron--positron
pair production; the optical depth of such a process is $\ge 1$ for $\ell
\ge 10$ (Svensson 1994). Provided the constraint on $R$ from the lowest
doubling time scale $R < 1.5 \times 10^{13} \ cm$ holds and the observed 2-10
keV luminosity $\sim 10^{42} \ erg \ s^{-1}$, $\ell_X \ge 2$ and $\ell_e$
of the injected electron distribution can be higher by an order of magnitude
if NGC4051 reproduces the multifrequency spectral behavior of NGC4151 (Yaqoob
1990). The condition for pair reprocessing could then also be achieved in
NGC4051.

The relation between power--law spectral photon index and physical parameters
in the simple case of an unsaturated thermal Comptonized model has been
derived analytically by Rybicki $\&$ Lightman (1979). $\Gamma$
depends inversely on the Compton parameter $y \equiv \frac{4 k T}{m_e c^2} \
\tau^2_{es}$. We could
therefore suppose the change in the observed [2--10 keV] spectral index
to be due to the gradual increase of $k T$ with decreasing flux.
However, if the soft photon input is constant, the pivot point
should yield an estimate of the energy of the soft photons
distribution, which is not compatible
with the observed value of $E_P \simeq 20 \ keV$ (see $\S$5.3).
Moreover, recent calculations by Chakrabarti $\&$ Titarchuk (1995) have
shown that spectral photon index as low as 1.5 can be only achieved
for a small part of the parameter space
with a thermal distribution of nonrelativistic electrons and $\Delta
\Gamma = 0.5$ requires variations of the accretion rate of two order
or magnitude. Such kind of dramatic flux changes has never been detected
in NGC4051, although they are not rare in Narrow Line Seyfert Galaxies.

A recent paper by Ghisellini $\&$ Haardt (1994)
demonstrated that it is possible
to set a one-to-one correspondence between $\Gamma$, $T$, $\ell_e$ and
$\ell_s$, provided that the thermal plasma is pair--dominated. The
{\it starvation ratio}
$\ell_e/\ell_s$ is fixed once the photon index is known and the $\ell_e$
spans more than 3 order of magnitude with the variation of the temperature
of plasma in pair equilibrium. Numerical simulations shows that
$\Delta \Gamma \sim 0.4$ requires a variation of the starvation ratio
$\sim 40$; that implies at least a factor of 10 of variation for the soft-energy
seed photon intensity. These numbers are in fairly good agreement with the
observed {\it ROSAT} variability (M$^c$Hardy {\it et. al.} 1995);
the upper limit on the central black hole mass is two order of
magnitude higher than that derived from the broadband {\it ASCA} variability (see $\S$
6.1) if the {\it ROSAT} time scale corresponds to the dynamical time scale of the
innermost allowed Schwarzschild orbit. However the maximum dynamical range
of the soft excess continuum component allowed by {\it ASCA} data is only $\sim$
2 ($1.0 \times 10^{-11} \le F_{bb} \le 2.0 \times 10{-11} \ erg \ s^{-1} \
cm^{-2}$ if the ``L'' model is applied to the spectra of the quartiles phases).
Therefore no evidence of the required dramatic flux variability of the soft
input component arises from the {\it ASCA} data; that casts some doubt on the
validity of such scenario for NGC4051.

An alternative explanation is provided by the non thermal Comptonization
models.
With the most simplifying assumptions, soft photons with a blackbody
distribution peaked at energy $x_s \equiv 2.8 k T/m_e c^2$ interact with
a monoenergetic distribution of relativistic electrons with Lorentz factor
$\gamma_0$. The compactness parameters $\ell_e$ and $\ell_s$ are defined as
above. A steady distribution of primary particles produces a spectral photon
index $\Gamma \sim 1.5$; if every scattered photon makes a pair with
equidistributed energies, the emerging spectrum of the second generation will
have $\Gamma \sim 1.75$. If pair saturation is achieved through successive
pair cascades, the value of the slope tends to $\Gamma \sim 2$, which is
consistent with the observed results and could explain the ``saturation''
effect shown by the $\Gamma_{2-10 \ keV}$ vs. $L$ relation in the present
analysis. Yaqoob (1992) studied in detail the physical condition
for the photon index variability observed in the Seyfert
1.5 galaxy NGC4151 to be achieved. The observed index--flux correlation in
NGC4051 can be reproduced by a variable Lorentz factor and a constant
electron injection
rate, with correlated input UV variations in order to leave $\ell_s/\ell_e$
constant. If the Lorentz factor
varies by a factor of $\sim 2$ it is constrained to lie between
$\sim 80 \left( \frac{3 \times 10^{-5}}{x_s} \right)^{1/2}$ and
$\sim 1100 \left( \frac{3 \times 10^{-5}}{x_s} \right)^{3/4}$.

Similar conclusions were derived in a recent paper by Torricelli-Ciamponi
$\&$ Courvoisier (1995), but in a slightly different scenario. In their
numerical model, both the thermal UV emission and the inverse Compton
reprocessing take place in the same hot plasma cloud. It need not to be
small in size nor optically thick to pair production and the UV
reprocessed photon luminosity is only a fraction of the observed. If
the observed X-ray continuum soft component is the hard tail of an
observed UV, there is therefore no need for any correlated variability
with the high energy ({\it i.e.} 2-10 keV) flux. The authors applied their
numerical model to the $F_{2-10 \ keV}$ vs. $\Gamma$ relation in Matsuoka
{\it et. al.} (1990) and derived a characteristic size of the cloud $\sim
10^{14} \ cm$ and a UV spectrum $kT \sim 30 \ eV$. Comparable results might be
expected if the fit were to include also the outlined {\it ASCA} results.
But the idea that the thermal and non-thermal radiation to be emitted in the
same physical region is not strongly supported by the current analysis,
regardless of the optical properties of the emitting plasma.
If we suppose an optically thin cloud with a numerical density from the Torricelli-Ciamponi $\&$
Courvoisier model ($n \sim 7 \times 10^{10} \ cm^{-3}$), its typical size
is $\sim 5 \times 10^{13} \ cm$; that implies $M_{bh}
\sim 10^8 M_{\odot}$, which is one order of magnitude higher than
the upper limit imposed by
the X-ray variability pattern (see $\S$6.1).
On the other hand, a slight trend of correlated
increase of the thermal emission with flux on timescales as low as several
$10^4 \ s$ might suggest to locate an optically thick emitting region at
$\sim 10^3 \ R_S$. In such a  case, the
characteristic temperature $T \simeq 180 \ eV$ is somewhat higher than
expected from the innermost region of an optically thick disk, but it is
fully consistent with the temperature required in a warm absorbed power law +
blackbody fit of a combined {\it ROSAT} and {\it Ginga} spectrum (Pounds
{\it et al.} 1994). Such outcome might be connected with the low luminosity and black hole mass, although it cannot rule out the effect
of Comptonization on the inner disk emission. Moreover it must be taken
into account that the {\it ASCA} low energy limited bandwidth alone is not
the most suitable for the exact determination of the temperature of any UV-soft X-ray component and the measured temperature is likely to be an upper limit
of the physical one.

We recall that an origin of the soft excess from a frequency-indipendent
scattering is ruled out by present analysis (see $\S$ 4.1).

The interpretative scenario is however complex and no unambiguous
interpretation can be claimed as a unique interpretation for the observational
data. The results from future broadband missions (like {\it XTE} or {\it SAX}) will
allow to determine with improved accuracy the parameters of all the spectral
components involved and to constrain the possible models and their degrees
of freedom.

\subsection{Iron line}

The {\it ASCA} AO2 observation of NGC4051 detected a clearly broad ($\sigma
\ge 0.2 \ keV$ at 90$\%$ level of confidence) and complex emission
line, that can be interpreted as a K-fluorescence iron line
whose profile has been modified by gravitational and Doppler shifts.
The profile and centroid of the line
allow to derive quantitative
constraints on the physical and geometrical properties of the emitting
region. The emission complex observed can be characterized, according to the equations
(1) to (3) in Matt {\it et al.} (1992a): $E_{c \alpha} = 6.09\pm0.03 \ keV$,
$\sigma_{\alpha} = 418 \pm 16 \ eV$ and $EW_{\alpha} = 320 \pm 30 \ eV$
(the results are the weighted mean of the parameters obtained applying the
above equatione to each detector continuum-subtracted spectrum). The emission
is therefore: a) significantly redshifted ($\sim 5 \%$) in comparision
to the expected energy for K-fluorescence emission line from neutral or
mildly ionized iron; b) broad; c) particularly
intense, being the average equivalent width of similar features observed
in other Seyfert 1 galaxies $\sim 150 \ eV$
(Nandra $\&$ Pounds 1994). Tanaka {\it et al.} (1995) reported the discovery
of an analogous complex emission feature in the 200,000 ks spectrum of
the Seyfert 1 galaxy MGC-6-30-15, giving the first conclusive evidence
of the effects due to the presence of a supermassive black hole in the
proximity of the
line radiation field. Explanations alternative to the relativistic
accretion disk seem to be excluded for that object (Fabian {\it et al.}
1995). Nonetheless the correct interpretation is far from clear and some
problems are left unsolved even if we consider only the NGC4051 results. Matt
{\it et al.} (1992) showed the $EW$ of a line emerging from an accretion disk
can hardly exceed $\sim 180 \ eV$, a value that is inconsistent with
the observed $EW \sim 270 \ eV$. 
An overabundance of iron of such order of magnitude could be a straightforward
explanation for such discrepancy.
Moreover, higher equivalent widths can be achieved
if the matter in the accretion disk is highly ionized (Zycki $\&$
Czerny 1994) due to the increased fluorescence yield and the reduced
photoelectric opacity;
$EW \sim 300 \ eV$ can be achieved if the mean ionization state of iron
is $\sim$ FeXXIII. In such case the rest energy of the emitted line is
expected to lie around $\sim 6.7 \ keV$.

The quoted results allow to put some significant constraints on the iron line
emission region. Following the fully relativistic treatment of Matt {\it
et al.} (1992a, 1992b), the observed centroid energy and intrinsic width
of the line profile require an inner radius $R_i \sim 10 \ R_S$, with
a comparable size of the nuclear source. The outer radius $R_o$ cannot be
tightly constrained, and might be as high as $\sim 10^3 \ R_S$ if $R_i$
could shrink to $\sim 6 R_S$. In case of a ionized disk, $EW \sim 300 \ eV$
can be obtained in the hypothesis of an extended source around a $\sim 10^6
M_{\odot}$ black hole, provided a typical size of the source $\sim 50 \ R_S$
and a sub-Eddington luminosity ($L/L_{Edd} \simeq 0.5$) are achieved. In
both cases the location of the innermost emitting region would lie in the
range $1 \div 5 \times 10^{12} \ cm$. Any response of the line to the
variation of the nuclear continuum should lag at most $\sim$ few hundreds
of seconds. Incidentally, we remark that we have not detected any evidence
of line variability in the
NGC4051 observations, while recent measures claiming correlated
iron-line/continuum variability in Seyfert 1 galaxies have recently appeared in
literature (Iwasawa {\it et al.} 1996, Yaqoob {\it et al.} 1996).

Alternatively, a contribution to the line emission
from a molecular torus surrounding the nucleus has been predicted by some
authors (Krolik $\&$ Kallmann 1987, Ghisellini, Haardt $\&$ Matt 1994,
Krolik, Madau $\&$ Zycki, 1994) and claimed as a possible explanation
for the narrow line observed in the Seyfert 1 galaxy NGC7469 (Guainazzi
{\it et al.} 1994, Leighly {\it et al.} 1995a). The line should be in such
case narrow and centred at the nominal K-fluorescence energy for neutral
iron ({\it i.e.} $\simeq 6.40 \ keV$).
This explanation would fit well
the multi-component structure of the iron line, that requires a narrow Gaussian
line at the nominal K$_{\alpha}$-fluorescence energy superimposed to the
relativistic double-horned profile, with $EW \simeq 60 \ eV$. Such an intensity
implies torus column density $N_H \sim$ few $10^{23} \ cm^{-2}$, although
much higher densities cannot be ruled out, since the $EW$ value is only loosely
constrained.

The discussion above is obviously affected by the choice of the self-consistent
continuum and line model outlined. A confirmation of the sketched scenario with
the upcoming broadband X-ray missions ({\it i.e. XTE} and
{\it SAX}) is therefore strongly needed in order to achieve a deeper understanding of
the physical process involved and of the relevant geometry.

\section{Conclusions}

The higher the resolution capabilities of the X--ray experiments, the
more interesting and complex the emission of AGNs appears.
The Seyfert 1 galaxy NGC4051
proves itself to be an outstanding target, that shares most of the peculiar
characteristics of the objects of its class, and namely:
\begin{itemize}
\item[a)] strong variability on time scales as short as few hundreds seconds.
\item[b)] Evidence of multicomponent X--ray spectrum, with an
apparent soft excess
emission superimposed on the higher energy power--law.
\item[c)] Detection of absorption edges from ionized oxygen species and
emission lines that represent clear evidence of warm absorbing matter along
the line of sight.
\item[d)] Variability constraints allow to locate the warm absorbing matter
at a distance from the nuclear source $\le 10^{17} \ cm$ with a numerical
density $n \ge 2 \times 10^{10} \ cm^{-3}$.
\item[e)] the power--law spectral photon index is intrinsically variable
(if reflection by a plane-parallel infinite disk is self-consistently supposed,
$\Delta \Gamma \sim 0.4$)
\item[f)] The iron line emission is redshifted, broad and very intense.
The outcomes can
be explained in terms of emission from a relativistic
accretion disk, either cold with moderate inclination ($\theta \sim
25^{\circ}$) or ionized and almost face-on. In the former case, the observed
EW is higher by a factor $\simeq 1.5 \div 2$ than the maximum allowed
by presently available solar--abundance relativistic models.
The puzzle of EW excess might be solved:
{\it i)}: postulating an {\it ad hoc} iron overabundance by a factor $\simeq 1.5$;
{\it ii)}: if the reflection is from an ionized face-on disk; {\it iii)}:
if a contribution to the iron emission complex comes from the putative
molecular torus surrounding the nucleus.
\end{itemize}
 
The evaluation of the above outlined results must take into account the
peculiar properties of NGC4051, which is a low-mass ($M_{bh} < 10^7 M_{\odot}$),
low-luminosity ($L_X \sim 10^{42} \ erg \ s^{-1}$) and strongly X-ray variable
(e-folding time $\sim$ few $10^2 \ s$) Seyfert 1. The mounting evidence for
a positive correlation between the 2-10 keV flux and photon index
point to an interpretation of the nuclear emission in terms
of inverse Compton reprocessing of UV seed photons by a non-thermal
distribution of relativistic electrons; the injected particles are likely
to be generated in the immediate proximity of the putative black hole,
while the exact location of the seed UV emitting region is still controversial.

NGC4051 exhibits all the main features often seen in Seyfert 1 X--ray spectra.
The many peculiar features of this source (variable warm absorber, complex
iron line structure, variable nonthermal continuum slope) allow a deeper
insight on the general properties of physical processes in AGNs.

Although a further effort is needed for a full understanding of the
complex interplay among the different spectral components and their
dynamics (the contribution of broadband experiments will be of the
primary importance to achieve such a goal), some interpretative
milestones can be set:
a non-thermal continuum, whose production
mechanism requires an efficient and continuous injection of a non-thermal
distribution of relativistic electrons, Compton reprocessing of the nuclear
radiation, a complex absorber with an intrinsic warm component within the
inner border of the BLR are the common elements any unified scenario
must cope with.

\section*{Acknowledgements}

Valuable discussions with Massimo Cappi and Karen Leighly improved greatly
the quality of this paper. The authors are deeply in debt with an anonymous
referee for her/his detailed revision. M.G. acknowledges support by M.U.R.S.T.
(Ministero dell'Universit\`a e della Ricerca Scientifica e Tecnologica).
\clearpage
\section*{References}
\noindent
Arnaud K.A. {\it et al.}, 1991, XSPEC User's Guide, ESA
TM-09 \\
Band D.L., Klein R.I., Castor J.I. $\&$ Nash J.K., 1990, Ap.J., 362, 90 \\
Cavaliere A. $\&$ Morrison P., 1980, Ap.J., 238, L63 \\
Chakrabarti $\&$ Titarchuk, 1995, Ap.J., 455, 623 \\
Comastri A., Setti G., Zamorani G. $\&$ Hasinger G., 1995, Astron $\&$
Astrophys., 296, 1 \\
Dickey J.M. $\&$ Lockman F.J., 1990, Ann. Rev. Astron. $\&$ Astrophys., 28,
215 \\
Edelson R., 1992, Ap.J., 401, 516 \\
Fabian A.C., Nandra K., Reynolds C.S., Brandt W.N., Otani C., Tanaka Y.,
Inoue H., Iwasawa K., 1995, Mon. Not. R. Astron. Soc., {\it in press} \\
Fabian A.C. $\&$ Rees M., 1979, in X-ray astronomy, Baity $\&$ Peterson
eds., p.381 \\
Fabian A.C., Rees M.J., Stella L. $\&$ White N.E., 1989,
Mon. Not. R. Astron. Soc.,
238, 729 \\
Fabian A.C., Kunieda H., Inoue S., Matsuoka M., Mihara T., Miyamoto S., Otani
C., Ricker G., Tanaka Y., Yamauchi M. $\&$ Yaqoob T., 1994, Publ. Astron.
Soc. Jap., 46, L59 \\
Ferland G.J., 1991, Ohio State University, Astronomy Departement Internal 
Report 91-01 \\
Fiore F., Elvis M., Mathur S., Wilkes B.J. $\&$ McDowell J.C., 1993, Ap.J.,
415, 129 \\
Fiore F., Perola G.C., Matsuoka M., Yamauchi M. $\&$ Piro L., 1992a, Astron.
$\&$ Astrophys., 262, 37 \\
Fiore F., Perola G.C., Yamauchi M. $\&$ Matsuoka M., 1992b, in Proc. 28th
Yamada Symposium on X--ray Astronomy, Nagoya, Japan, Tanaka Y. $\&$ Koyama K.
eds., Tokyo:Universal Academy Press \\
Gendreau K.C., Mushotsky R., Fabian A.C., Holt S.S., Kii T., Serlemitos P.J.,
Ogasaka Y., Tanaka Y. {\it et al.},
1995, Publ. Astron. Soc. Jap., 47, L5 \\
George I.M. $\&$ Fabian A.C., 1991, Mon. Not. R. Astr. Soc., 249, 352 \\
George I.M., Turner T.J. $\&$ Netzer H., 1995, Ap. J. Lett., 438, L67 \\      
Ghisellini G. $\&$ Haardt F., 1994, Ap.J.Lett, 429, L53 \\
Ghisellini G., Haardt F. $\&$ Matt G., 1994, Mon. Not. R. Astron. Soc.,
267, 743 \\
Guainazzi M., Matsuoka M., Piro L., Mihara T. $\&$ Yamauchi M., 1994,
Ap.J.Lett., 436, L35 \\
Guilbert P.W., Fabian A.C. $\&$ Rees M., 1983, Mon. Not. R. Astr. Soc., 205,
593 \\
Guilbert P.W. $\&$ Rees M.J., 1988, Mon. Not. R. Astr. Soc., 233, 475 \\
Huchra J. $\&$ Burg R., 1992, Ap.J., 393, 90 \\
Iwasawa K., Fabian A.C., Reynolds C.S., Nandra K., Otani C., Inoue H., Hayashida H., Brandt W.M., Dotani T., Kunieda H., Matsuoka M. $\&$ Tanaka Y., 1996, Mon. Not. R. Astron. Soc., {\it in press} \\
Krolik J.H., $\&$ Kallmann T.R., 1987, Ap.J.Lett., 320, L5 \\
Krolik J.H., Kriss G.A., 1995, Ap.J., 447, 512 \\
Krolik J.H., Madau P. $\&$ Zycki P., 1994, Ap.J.Lett., 420, L57 \\
Krolik J.H., McKee, Tarter, 1981, Ap.J., 249, 422 \\
Kunieda H., Hayakawa S., Tawara Y., Koyama K., Tsuruta S., Leighly K.,
1992, Ap.J., 384, 482 \\
Lampton M., Margon B. $\&$ Bowyer S., 1976, Ap. J., 208, 177 \\
Lawrence A., Watson M.G., Pounds K.A., Elvis M., 1985, Mon. Not. R. Astron.
Soc., 217, 685 \\
Leighly K., Kunieda H., Awaki H., Tsuruta S., 1995, {\it submitted to
Ap.J.} \\
Leighly K., Mushotzky R., Yaqoob T., Kunieda H. $\&$ Edelson R., 1996,
{\it in press} \\
Liedhal D.A., Kahn S.M., Osterheld A.L. $\&$ Goldstein W.H., 1990, Ap.J.,
350, L37 \\
Lightman A.P. $\&$ White T.R., 1988, Ap.J., 335, 57 \\
Marshall F.E., Holt S.S., Mushotzky R.F., Becker R.H., 1983, Ap.J.Lett., 269,
L31 \\
Matsuoka M. Yamauchi M. $\&$ Piro L., 1989, in Proc. 23rd ESLAB Symposium, 
Hunt J. $\&$ Battrick B. eds., ESA SP-296, pag.985 \\
Matsuoka M., Piro L., Yamauchi M., Murakami T., 1990, Ap.J., 361, 440 \\
Matt G., Perola G.C. $\&$ Piro L., 1991, Astron. $\&$ Astrophys., 247, 25 \\
Matt G., Perola G.C., Piro L. $\&$ Stella L., 1992a, Astron. $\&$ Astrophys.,
257, 63 \\
Matt G., Perola G.C., Piro L. $\&$ Stella L., 1992b, Astron. $\&$ Astrophys.,
263, 453 \\
M$^c$Hardy I.M., Green A.R., Done C., Punchnarewicz E.M., Mason K.O.,
Branduardi--Raymond G. $\&$ Jones M.H., 1995, Mon. Not. R. Astron. Soc.,
273, 549 \\
Mihara T., Matusoka M., Mushotzky R.F., Kunieda H., Otani C., Miyamoto S.,
Yamauchi M., 1994, Pub. Astron. Soc. Jap., 46, L137 (Paper I)\\
Mushotzky R.F., Fabian A.C., Iwasawa K., Kunieda H., Matsuoka M., Nandra K.,
$\&$ Tanaka Y., 1995, Mon. Not. R. Astron. Soc., 272, L9 \\
Nandra K. $\&$ Pounds K.A., 1994, Mon. Not. R. Astron. Soc., 268, 405 \\
Netzer H., 1993, Ap.J., 411, 594 \\
Netzer H., Turner T.J., George I.M., 1994, Ap.J., 435, 106 \\
Osterbrock D.E., Martel A., 1993, Ap.J., 414, 552 \\
Otani C., 1996, Ph.D. thesis, Tokyo University \\
Otani C., Kii T., Reynolds C.S., Fabian A.C., Iwasawa K., Hayashida K., Inoue H., Kunieda H., Makino F., Matusoka M., Tanaka Y., 1996, Publ. Astron. Soc. Jap., 48, 211 \\
Otani C., $\&$ Dotani T., 1994, Asca News n.2, 25 \\
Pounds K.A., Nandra K., Fink H.H., Makino F., 1994, Mon. Not. R. Astron.
Soc., 267, 193 \\
Ptak A., Yaqoob T., Serlemitos P.J. $\&$ Mushotzky R.F., 1994, Ap. J. Lett.,
436, L31 \\    
Reynolds C.S. $\&$ Fabian A.C., 1995, Mon. Not. R. Astron. Soc., 273, 1167 \\
Reynolds C.S., Fabian A.C., Nandra K., Inoue H., Kunieda H. $\&$ Iwasawa K.,
1995, Mon. Not. R. Astron. Soc., {\it in press} \\
Rybicki G.B. $\&$ Lightman A.P., 1979, Radiative processes in astrophysics,
New York:John Wiley $\&$ Sons \\
Svensson R., 1994, Ap.J. Suppl. Ser., 92, 585 \\
Tanaka Y., Inoue H., $\&$ Holt S.S., 1994, Publ. Astron. Soc. Jap., 46, L37 \\
Tanaka Y., Nandra K., Fabian A.C., Inoue H., Otani C., Dotani T., Hayashida K.,
Iwasawa K. {\it et al.}, 1995, Nature,
375, 659 \\
Torricelli-Ciamponi G. $\&$ Courvoisier T. J.-L., 1995, Astron. $\&$
Astrophys., 296, 651 \\
Turner T.J., Weaver K.A., Mushotzky R.F., Holt S.S. $\&$ Madjeski G.M.,
1991. Ap.J., 381, 85 \\
Urry C.M., Arnaud K., Edelson R.A., Kruper J.S. $\&$ Mushotzky R.F., 1989, in
AGN and X--ray background, Proc.32rd ESLAB Symposium, Vol.2, Hunt J. $\&$
Battrick B. eds., ESA, SP-296, p.789 \\
Yaqoob T., 1990, {\it Ph.D. thesis}, University of Leicester \\
Yaqoob T., 1992, Mon. Not. R. Astron. Soc., 258, 198 \\
Yaqoob T., Edelson R., Weaver K.W., Warwick R., Mushotzky R.F. Serlemitos
P.J. $\&$ Holt S.S., 1995, Ap. J., 453, 81 \\
Yaqoob T., Serlemitos P.J., Turner T.J., George I.M., Nandra K., 1996,
Ap.J.Lett., {\it in press} \\
Yaqoob T. $\&$ Warwick R., 1991, Mon. Not. R. Astron. Soc., 248, 773 \\
Yaqoob T., Warwick R., Makino F. {\it et al.}, 1993 Mon. Not R. Astron. Soc.,
262, 435 \\
Zycki P.T. $\&$ Czerny B., 1994, Mon. Not. R. Astron. Soc., 266, 653 \\
\clearpage
\section*{Figure captions}
\noindent
{\bf Figure 1.} Broadband SIS0 and GIS2 light curves with binning time
$\Delta t = 312.5 \ s$. Time is expressed in seconds starting from the
beginning of the observation. \\[0.1cm]
\noindent
{\bf Figure 2.} Distribution function of the lowest doubling times for
the broad band curves of Figure 1 ($\Delta t = 100 \ s$). The distribution
functions have been fit with an empirical equation of the form: $A[1 -
exp(-\Delta t/\tau)]$. Best-fit values are: $\tau_{SIS0} = 500 \pm 300 \ s$
and $\tau_{GIS2} = 400 \pm 300 \ s$. \\[0.1cm]
\noindent
{\bf Figure 3.} Broadband spectra ({\it upper panels}) and residuals
in units of $\sigma$
({\it lower panels}) when a simple power--law with cold absorption is
applied. The energy channels have been rebinned in order to have a S/N
ratio $\ge 25$.
Soft excess below $E \sim 1 \ keV$ and
a broad emission line feature around $E \sim 6.1 \ keV$ are clearly visible.
\\[0.1cm]
\noindent
{\bf Figure 4.} Data/model ratio when a {\sc Cloudy} ``warm absorber''
model is applied to the SIS0+1 data. Bad modelling of the OVII edge and a
prominent emission--like feature at energy $E \sim 0.95 \ keV$ are clearly
seen. \\[0.1 cm]
\noindent
{\bf Figure 5.} Data/model ratio for a simple power-law
model applied to the SIS0 and SIS1 spectrum in the energy range [3-10 keV].
The energy channel have been rebinned in order to have a S/N ratio $\ge 15$.
In the upper panel a centroid energy vs. Gaussian dispersion contour plot
is shown for the emission complex main component when the power-law +
flat reflection + emission complex model is applied to the spectra of
all detectors simultaneously. The outermost contour corresponds to 90$\%$
level of confidence for two interesting parameters ($\Delta \chi^2 = 4.61$).
At such level of confidence the line is broad ($\sigma \ge 0.2 \ keV$). \\
\noindent
{\bf Figure 6.} PHA ratio between the HS and LS phase SIS0 spectra (definition
of intensity phases is given in text). \\[0.1 cm]
\noindent
{\bf Figure 7.} [2.5-5.0 keV] and
[0.7-1.3 keV] bands and SR SIS0+1 light curves for a binning
time $\Delta t = 500 \ s$. \\[0.1 cm]
\noindent
{\bf Figure 8.} SR vs. [0.7-5 keV] count rate for the SIS0+1 light curve
with binning time $\Delta t = 5000 \ s$. The
reduced $\chi^2$ for constant hypothesis is 8.9 and the linear correlation
coefficient for 29 degrees of freedom $r_{29} = 0.71 \pm 0.07$. \\[0.1 cm]
\noindent
{\bf Figure 9.} OVII edge energy vs. optical depth contour plot for High State
(HS), Low State subtracted of the 10th orbit (LS$^{\star}$) and 10th orbit.
The 10th orbit coincides with the $\sim 2 \times 10^4 \ s$ interval
when the source flux is minimum.
The outermost contour corresponds to the 99$\%$ level of confidence for
two interesting parameters ($\Delta \chi^2 = 9.21$). The definition of
the intensity phases is given in text. \\[0.1 cm]
\noindent
{\bf Figure 10.} The same as Figure 10 for OVIII. \\[0.1 cm]
\noindent
{\bf Figure 11.} The same as Figure 9 when a model similar to Model ``L'' is
applied but a bremsstrahlung ({\it left panel}) or power-law ({\it right
panel}) emission model is substituted for the blackbody. \\[0.1cm]
\noindent
{\bf Figure 12.} Photon index vs. [2--10 keV] luminosity for both AO2 and
PV phase data (indicated by an arrow). \\[0.1 cm]
\end{document}


\begin{center}
{\sc Table 2}: Best fit parameters for
low energy ($E \le 5 \ keV$) SIS0+1 spectral fits
(PL = power-law, BB = blackbody, ED = absorption edge, GA = gaussian line,
BREMS = bremsstrahlung, DISKBB = accretion disk multitemperature blackbody)
 \\[0.25 cm]
\end{center}
\begin{tabular}{lcccccccc} \hline \hline
Model 	& $F^{\dag}_{se}$ 	& $E_{OVII} \ (keV)$	& $\tau_{OVII}$	& $E_{OVIII} \ (keV)$	& $\tau_{OVIII}$	& $E_3 \ (keV)$	& $\tau_3$ or $EW_3 \ (eV)$	& $\chi^2/$d.o.f.	\\ \hline 
PL + BB	& $0.61^{+0.11}_{-0.10}$	& ...	& ...	& ...	& ...	& ...	& ...	& 768/604	\\
PL + BB + 1 ED	& $0.71^{+0.08}_{-0.09}$	& $0.752\pm0.016$	& $0.23\pm0.04$	& ...	& ...	& ...	& ...	& 718/602	\\
PL + BB + 2 EDs	& $0.94^{+0.12}_{-0.09}$	& $0.74\pm0.01$	& $0.36\pm0.04$	& $0.92\pm0.02$	& $0.22\pm0.04$	& ...	& ...	& 634/600 \\
PL + BB + 3 EDs (E)$^{\ddag}$	& $1.1\pm0.6$	& $0.733^{+0.014}_{-0.095}$	& $0.35^{+0.04}_{-0.13}$	& $0.86^{+0.04}_{-0.03}$	& $0.12^{+0.05}_{-0.06}$	& $0.96\pm0.02$	& $0.12^{+0.10}_{-0.02}$	& 624/598	\\
PL + BB + 2EDs + GA (L)$^{\ddag}$	& $1.00^{+0.15}_{-0.13}$	& $0.734\pm0.013$	& $0.35^{+0.04}_{-0.03}$	& $0.89\pm0.02$	& $0.31^{+0.05}_{-0.06}$	& $0.94\pm0.02$	& $10^{+3}_{-5}$	& 624/598 \\
PL + BREMS +  2ED + GA	& $1.8^{+1.2}_{-1.0}$	& $0.737\pm0.013$	& $0.29\pm0.05$	& $0.90\pm0.02$	& $0.22^{+0.06}_{-0.04}$	& $0.93^{+0.07}_{-0.02}$	& $8^{+6}_{-7}$	& 635/598	\\
PL + DISKBB + 2ED + GA	& $1.2^{+1.1}_{-0.6}$	& $0.734\pm0.012$	& $0.31^{+0.05}_{-0.06}$	& $0.90\pm0.02$	& $0.26\pm0.05$	& $0.93\pm0.02$	& $9\pm5$	& 626/598	\\
PL + PL + 2ED + GA	& $3.2^{+0.9}_{-0.8}$	& $0.747^{+0.014}_{-0.016}$	& $0.21\pm0.06$	& $0.92^{+0.050}_{-0.013}$	& $0.15^{+0.06}_{-0.04}$	& $0.94\pm0.10$	& $<20$	& 669/598 \\ \hline
\end{tabular}
$^{\dag}$flux of the soft excess component in the [0.4-2 keV] band expressed in units of $10^{-11} \ erg \ s^{-1} \ cm^{-2}$ \\
$^{\ddag}$referred to as ``Model E'' and ``Model L'' in text respectively
\clearpage
\begin{center}
{\sc Table 3}: Best--fit parameters and $\chi^2$ values for 
High energy ($E \ge 2.3 \ keV$) SIS0+1 and GIS2+3
simultaneous spectral fits \\[0.25 cm]
\end{center}
\begin{tabular}{lccccccccc} \hline \hline
Model	& $\Gamma$	& $E_1 \ (keV)$	& $\sigma_1 \ (eV)$	& $EW_1 \ (eV)$	& $E_2 \ (keV)$	& $EW_2 \ (eV)$	& $E_3 \ (keV)$	& $EW_3 \ (keV)$	& $\chi^2/$d.o.f.	\\ \hline 
\multicolumn{10}{l}{SIS0+1} \\ 
PL + 1GA	& $1.84\pm0.06$	& $6.3\pm0.2$	& $0.5^{+0.3}_{-0.2}$	& $350^{+170}_{-150}$	& ...	& ...	& ...	& ...	& 553/552 \\
PL + 2GA	& $1.85\pm0.07$	& $6.2\pm0.2$	& $0.7\pm0.4$	& $320\pm190$	& $6.44^{+0.05}_{-0.24}$ & $40\pm30$	& ...	& ...	& 546/550 \\
PL + 3GA	& $1.84^{+0.07}_{-0.04}$	& $6.3\pm0.2$	& $0.5^{+0.4}_{-0.2}$	& $270^{+190}_{-120}$	& $6.44^{+0.05}_{-0.26}$	& $40\pm30$	&	$5.4^{+0.7}_{-0.5}$	& $21\pm17$	& 542/548 \\	\hline
\multicolumn{10}{l}{GIS2+3} \\  
PL + 1GA	& $1.89^{+0.03}_{-0.06}$	& $6.0\pm0.3$	& $0.9 \pm 0.6$	& $500^{+600}_{-300}$	& ...	& ...	& ...	& ...	& 607/674	\\
PL + 2GA	& $1.85\pm0.05$	& $6.3\pm0.2$	& $0.5 (fixed)$	& $250 \pm 80$	& ... & ... & $5.5 \pm 0.7$ & $<60$ & 617/672 \\
PL + 3GA	& $1.83^{+0.05}_{-0.03}$	& $6.41 \pm 0.08$	& $0 (fixed)$	& $160^{+50}_{-40}$	& ... & ... & $5.62 \pm 0.12$	& $60^{+30}_{-20}$	& 606/670  \\  	\hline
\end{tabular}
\clearpage
\begin{center}
{\sc Table 5}: Log of temporal subdivision of the observation, SIS1 count rate
(mean count rate = 1.11 cts/s), OVII and OVIII absorption edges optical
depths and [2--10 keV] photon index in the E model case
\\[0.25cm]
\end{center}
\begin{tabular}{lcccccc} \hline \hline
Orbit	& $t_{start}-t_{end} \ (Ks)^{\dag}$	& Exposure (s)	& $CR \ s^{-1}$	& $\tau_{OVII}$	& $\tau_{OVIII}$	& $\Gamma_{2.3-10 \ keV}$	\\ \hline 
1	& 0-20	& 10448	& 1.60	& $0.28\pm0.08$	& $0.16\pm0.07$	& $2.08\pm0.09$	\\
2	& 20-60 & 15105	& 0.90	& $0.37\pm0.09$	& $0.17\pm0.08$	& $1.68^{+0.05}_{-0.06}$	\\
3	& 60-64	& 1382	& 1.48	& $< 0.57$	& $<0.35$	& $2.0\pm0.2$	\\
4	& 64-75	& 5291	& 1.14	& $0.48^{+0.15}_{-0.11}$	& $0.22\pm0.12$	& $1.76\pm0.09$	\\
5	& 75-81.5	& 3521	& 1.28	& $0.23^{+0.16}_{-0.15}$	& $0.18\pm0.14$	& $1.83^{+0.54}_{-0.12}$	\\
6	& 81.5-88.5	& 3884	& 0.96	& $<0.36$	& $<0.26$	& $1.82\pm0.11$	\\
7	& 88.5-95	& 3231	& 1.22	& $0.30^{+0.22}_{-0.19}$	& $0.16\pm0.16$	& $1.98^{+0.16}_{-0.13}$	\\
8	& 95-108	& 5281	& 0.81	& $0.32\pm0.16$	& $<0.18$	& $1.81\pm0.03$	\\
9	& 108-115	& 2499	& 1.31	& $0.34\pm0.22$	& $0.21^{+0.17}_{-0.16}$	& $2.1\pm0.3$	\\
10	& 115-140	& 10289	& 0.66	& $0.63^{+0.14}_{-0.13}$ & $<0.16$	& $1.6\pm0.4$	\\
11	& 140-155	& 6646	& 1.50	& $0.35\pm0.11$ & $<0.13$	& $1.91^{+0.07}_{-0.06}$ \\ \hline
\end{tabular}
\\
$^{\dag}$times are calculated from the start of the observation
\clearpage
\begin{center}
{\sc Table 6}: Best--fit parameters for the
low energy spectral features in
SIS0+1 spectra of time--resolved phases \\[0.25cm]
\end{center}
\begin{tabular}{lccccccc} \hline \hline
Phase	& $E_{OVII} \ (keV)$	& $\tau_{OVII}$	& $E_{OVIII} \ (keV)$	& $\tau_{OVIII}$	& $E_3 \ (keV)$	& $\tau_3$ or $EW_3 \ (eV)$	& $\chi^2_{\nu}$	\\ \hline
\multicolumn{8}{l}{Model L} \\
High State	&  $0.726\pm0.012$	& $0.29\pm0.07$	& $0.88\pm0.04$	& $0.30\pm0.06$	& $0.94^{+0.02}_{-0.06}$	& $8^{+9}_{-6}$	& 1.06	\\
Low State	&  $0.739^{+0.013}_{-0.008}$	& $0.44^{+0.08}_{-0.06}$	& $0.90^{+0.05}_{-0.02}$	& $0.31^{+0.09}_{-0.11}$	& $0.93^{+0.02}_{-0.03}$	& $< 14$	& 1.06	\\
10th orbit	&  $0.722\pm0.017$	& $0.66\pm0.12$	& $0.91^{+0.02}_{-0.03}$	& $0.44\pm0.10$	& $0.94\pm0.08$	& $5^{\dag}$	& 1.01 \\
Low State$^{\star}$	&  $0.743\pm0.013$	&$0.39^{+0.09}_{-0.10}$	& $0.87\pm0.04$	& $0.34^{+0.10}_{-0.17}$	& $0.90^{+0.02}_{-0.05}$	& $16^{+10}_{-8}$	& 0.99	\\ \hline
\multicolumn{8}{l}{Model E} \\ 
High State	&  $0.723\pm0.013$	& $0.29^{+0.06}_{-0.05}$	& $0.86^{\dag}$	& $0.14\pm0.07$	& $0.96\pm0.03$	& $0.14\pm0.05$	& 1.06 \\
Low State	&  $0.738^{+0.015}_{-0.007}$	& $0.45\pm0.08$	& $0.86^{\dag}$	& $<0.13$	& $0.945^{+0.02}_{-0.03}$	& $0.21^{+0.09}_{-0.05}$	& 1.06 \\
10th orbit	&  $0.722^{+0.017}_{-0.010}$	& $0.66\pm0.12$	& $0.86^{\dag}$	& $<0.17$	& $0.96^{+0.01}_{-0.05}$	& $0.38^{+0.11}_{-0.14}$	& 1.01	\\
Low State$^{\star}$	&  $0.746^{+0.012}_{-0.013}$	& $0.42^{+0.11}_{-0.12}$	& $0.86^{\dag}$	& $<0.20$	& $0.94\pm0.03$	& $0.21^{+0.09}_{-0.08}$	& 0.96	\\ \hline
\end{tabular}
$^{\dag}$ unconstrainted
\clearpage
\begin{center}
{\sc Table 1}: - Good time interval selection criteria \\[0.25cm]
\begin{tabular}{cc} \hline \hline
SIS0+1	& GIS2+3	\\ \hline
\multicolumn{2}{c}{SAA.eq.0} \\
\multicolumn{2}{c}{T$\_$SAA.gt.60} \\
\multicolumn{2}{c}{BR$\_$EARTH.gt.20} \\
\multicolumn{2}{c}{ELV$\_$MIN.gt.5} \\
\multicolumn{2}{c}{T$\_$DY$\_$NT.gt.5} \\
COR$\_$MIN.gt.6	& COR$\_$MIN.gt.7 \\
SX$\_$PIXLY.lt.400$^{\dag}$	& GZ$\_$L1.gt.0$^{\ddag}$	\\ \hline
\end{tabular}
\\
$^{\dag}$X=0,1, Y=1,3 \\
$^{\ddag}$Z=2,3 \\[0.25cm]
SAA = South Atlantic Anomaly (0 = passage excluded) \\
T$\_$SAA = Time after passage through SAA (in s) \\
BR$\_$EARTH = Angle between pointing direction and Earth's terminator (in
degrees) \\
ELV$\_$MIN = Angle between pointing direction and Earth's limb (in degrees) \\
T$\_$DY$\_$NT = Time after day/night transition (in s) \\
COR$\_$MIN = Minimum cut--off rigidity allowed (in GeV/$c$) \\
SX$\_$PIXLY = Maximum number of counts in chip Y of detector SISX \\
GX$\_$L1 = GIS2 and GIS3 discriminator \\
\end{center}
\clearpage
\begin{center}
{\sc Table 7}: Best--fit parameters for SIS0+1 and GIS2+3
simultaneous high energy fits on AO2 data quartiles with constant
direct vs. reflected normalizations ratio
(quartile's definition
in text,
in brackets the [2--10 keV] luminosity in units $10^{42} \ erg \ s^{-1}$)
\\[0.25cm]
\end{center}
\begin{tabular}{lcccccccc} \hline \hline
Phase	& $\Gamma_{2.3-10 \ keV}$	& $E^{Fe}_{line} \ (keV)^{\ddag}$	& $\sigma^{Fe}_{line} \ (keV)^{\ddag}$	& $EW^{Fe}_{line} \ (eV)^{\ddag}$	& $\Gamma_{5-10 \ keV}/\Gamma_{2.3-5 \ keV}$	& $\chi^2$/d.o.f.	& F-test$^{\dag}$	& $\Omega/2 \pi^{\dag}$	\\ \hline 
1st quartile (0.33)	& $1.57\pm0.08$	& $6.28\pm0.14$	& $0.20^{+0.12}_{-0.20}$	& $250^{+90}_{-100}$	& $1.19\pm0.19$		& 295/265	& 2.4	& $0(<3.2)$	\\
2nd quartile (0.56)	& $1.72\pm0.07$	& $6.58\pm0.14$	& $0.08^{+0.17}_{-0.08}$	& $180^{+100}_{-90}$	& $1.08^{+0.16}_{-0.20}$	& 299/294	& 0.1	& $0(<4.9)$	\\
3rd quartile (0.70)	& $1.88\pm0.07$	& $6.17^{+0.13}_{-0.46}$	& $0.13^{+0.69}_{-0.13}$	& $200\pm90$	& $0.93^{+0.15}_{-0.18}$	& 301/344	& 0.3	& $2^{+5}_{-2}$	\\
4th quartile (1.13)	& $2.02^{+0.07}_{-0.05}$	& $6.30\pm0.10$	& $0.4^{+0.5}_{-0.3}$	& $270\pm120$	& $0.88^{+0.09}_{-0.13}$	& 578/581 	& 2.5	& $4\pm3$	\\ \hline
\end{tabular}
\\
$^{\dag}$ the quoted F-test is calculated after allowing the $\Omega/2 \pi$
parameter to float in the reflection model. The resulting best-fit values for this
parameter are listed in the adjacent column \\[0.1cm]
$^{\ddag}E$ and $\sigma$ are referred to the main component of the emission
complex; the $EW$ instead is referred to the complex as a whole \\
\clearpage
\begin{center}
{\sc Table 8}: Best--fit parameters for SIS0+1 and GIS2+3
simultaneous high energy fits on AO2 data quartiles with reflected
component normalization inversely correlated with total flux
\\[0.25cm]
\end{center}
\begin{tabular}{lcccccccc} \hline \hline
Phase	& $\Gamma_{2.3-10 \ keV}$	& $E^{Fe}_{line} \ (keV)^{\ddag}$	&  $EW^{Fe}_{line} \ (eV)^{\ddag}$	&  $\chi^2$/d.o.f.	&  $\Omega/2 \pi$ (fixed)	\\ \hline 
1st quartile 	& $1.66\pm0.09$	& $6.0\pm0.5$	&  $100^{+130}_{-100}$	& 288/277 & 1.72 \\
2nd quartile 	& $1.79^{+0.22}_{-0.10}$	& $6.6\pm0.4$	& $110^{+180}_{-110}$	& 290/277	& 1.2 \\
3rd quartile 	& $1.88\pm0.07$	& $6.2\pm0.2$	& $200\pm90$	& 322/277 & 1 \\
4th quartile 	& $2.05^{+0.09}_{-0.06}$	& $5.8\pm0.5$	& $150\pm100$	& 290/277 	& 0.77 \\ \hline
\end{tabular}
\\
$^{\ddag}$line parameters refers to the main component of the emission
complex \\
\clearpage
\begin{center}
{\sc Table 4}: Best--fit parameters for the line emission from a
relativistic accretion disk model \\[0.25 cm]
\begin{tabular}{cccccc} \hline \hline
$q$	& $R_i/R_S$	& $R_o/R_S$	& $\theta$	& $EW \ (eV)$	& $\chi^2$/d.o.f. \\
\hline 
\multicolumn{6}{l}{$E_{\alpha} \equiv 6.4 \ keV$ (cold disk)} \\
$-1^{\dag}$	& $9^{+\infty}_{-6}$	& $28^{+7}_{-5}$	& $25^{+3}_{-2}$
	& $270\pm40$	& 1157/1230 \\
$-3^{\dag}$	& $10^{+7}_{-4}$	& $34^{+\infty}_{-9}$ & $28^{+4}_{-6}$	& $270^{+30}_{-60}$	&	1157/1230	\\
$-2\pm3$	& $10^{\dag}$	& $30^{\dag}$	& $27\pm4$	& $270\pm50$	
& 1157/1230	\\ \hline
\multicolumn{6}{l}{$E_{\alpha} \equiv 6.7 \ keV$ (ionized disk)}	\\
$-3^{\dag}$	& $7\pm3$	& 5000$^{\ddag}$	& $10^{+6}_{-10}$
	& $260^{+40}_{-50}$	& 1157/1230	\\
$-2.9\pm0.2$	& $8^{\dag}$	& $5000^{\dag}$	& $10^{+6}_{-10}$	& 
$260\pm50$	& 1157/1230 \\ \hline
\end{tabular}
\\
$^{\dag}$ fixed \\
$^{\ddag}$ unconstrained
\end{center}